# A Unified Description of Electron-Phonon Coupling and Ion Migration in Metal Halide Perovskites

Bo Cai[1,+,*], Yan Yang[1,+], Yoshiki Sugai[2], Maddison Wiles[3], Dongxu He[4], Yang Yang[1], Junmin Xia[1,*], Shufen Chen[1,*], Carla Verdi[3,*], Siyu Chen[5,6,7,*], Nan Zhang[8], Ming-Gang Ju[9], Chao Liang[8,*], and Julian A. Steele[2,3,*]

*1 State Key Laboratory of Flexible Electronics (LoFE) & Institute of Advanced Materials (IAM), Nanjing University of Posts and Telecommunications, Nanjing, 210023, P. R. China.*

*2 Australian Institute for Bioengineering and Nanotechnology, The University of Queensland, Brisbane, QLD, 4072 Australia.*

*3 School of Mathematics and Physics, The University of Queensland, Brisbane, QLD, 4072 Australia.*

*4 School of Chemical Engineering, The University of Queensland, Brisbane, QLD, 4072 Australia.*

*5 TCM Group, Cavendish Laboratory, University of Cambridge, Cambridge CB3 0HE, UK*

*6 Department of Materials Science and Metallurgy, University of Cambridge, Cambridge CB3 0FS, UK*

*7 European Theoretical Spectroscopy Facility, Institute of Condensed Matter and Nanosciences, Université catholique de Louvain, Louvain-la-Neuve, 1348, Belgium*

*8 State Key Laboratory of Electrical Insulation and Power Equipment, MOE Key Laboratory for Nonequilibrium Synthesis and Modulation of Condensed Matter, School of Physics, Xi'an Jiaotong University, Xi'an, 710049, P. R. China.*

*9 Key Laboratory of Quantum Materials and Devices of Ministry of Education, School of Physics, Southeast University, 211189 Nanjing, China.*

Email: iambcai@njupt.edu.cn; iamjmxia@njupt.edu.cn; iamsfchen@njupt.edu.cn; c.verdi@uq.edu.au; sc2090@cam.ac.uk; chaoliang@xjtu.edu.cn; julian.steele@uq.edu.au.

## Abstract

The remarkable optoelectronic properties of metal halide perovskites are closely linked to their unusually soft and polar chemical bonds that enable both strong electron–phonon interactions and ion migration. Yet these two defining characteristics have largely been treated as independent consequences of the same underlying chemical bonding. Here we show that they originate from a common electronic-structure framework by developing a general description linking lattice dynamics, electron–phonon coupling, and halide ion migration across representative Pb-based, Sn-based, and double perovskites. Spectrally resolved phonon-mode contributions demonstrate that the low-frequency shearing modes dominate halide migration, whereas high-frequency stretching modes govern carrier scattering through the Fröhlich interaction in all three compositions. We introduce an orbital hybridization descriptor to unify these findings, which connects metal–halide bonding characteristics with the migration barrier energies and Fröhlich coupling strengths, indicating a cooperative evolution of these two properties. These findings provide a generalized microscopic mechanism for simultaneously optimizing charge and ionic transport in soft semiconductors.

## Introduction

Metal halide perovskite (MHP) semiconductors have transformed optoelectronics research in recent years as they combine exceptional photovoltaic and light-emitting performance with cheap and scalable fabrication(*1*–*5*). Unlike conventional semiconductors, however, their ease of production stems from an intrinsically soft crystal structure with low crystal formation energies(*6*, *7*). MHPs consequently exhibit dynamically disordered local structure(*8*) and strong lattice anharmonicity(*9*–*11*), yielding a metal-halide sub-lattice that challenges the static lattice approximation commonly applied to inorganic semiconductors. For example, large-amplitude atomic motions generate dynamic lattice distortions and local electric fields that couple strongly to charge carriers(*8*, *12*) while simultaneously lowering energy barriers for ionic migration in the bulk(*13*, *14*).

It follows that many of the defining properties of MHPs arise from a complex interplay between structural dynamics and electronic behavior(*1*). Strong electron–phonon coupling, for example, governs charge-carrier scattering and mobility, hot-carrier cooling, band-gap renormalization, and optical linewidth broadening (i.e., color purity)(*15*–*18*), while also promoting the formation of large polarons that screen carriers from defects and contribute to their unusually long lifetimes and diffusion lengths(*12*, *18*, *19*). Concurrently, ion migration via the movement of lattice vacancies influences device stability, hysteresis, phase segregation, and operational degradation(*13*, *14*, *20*, *21*).

To date, studies connecting charge transport and ion migration have largely focused on changes in optoelectronic properties following phase segregation or on the role of photo-induced polarons in facilitating ionic migration(*22*–*25*). Notably, the majority of research into these dynamic properties has progressed along independent trajectories(*21*, *26*–*28*). In the field of electron–phonon coupling (EPC), a consensus has emerged that the strong Fröhlich interaction associated with longitudinal optical (LO) phonons dominate carrier scattering at room temperature(*12*, *19*). These dynamic polar fluctuations originate primarily from the metal–halide sub-lattice(*29*), whose magnitude is typically 3–30 times larger than that of the caged organic A-site cations and controls carrier protection and defect tolerance(*8*, *29*–*31*). On the other hand, ion migration is understood to be governed by anharmonic lattice fluctuations and vacancy-assisted hopping. The migration energy barriers of halide anions are typically small, falling within the range of 0.1~0.6 eV(*13*, *20*, *21*, *32*, *33*), and reflect the weak restoring forces that confine ions around their equilibrium sites, in line with the characteristic low-energy phonons present in the soft metal–halide framework.

An overview of these lattice dynamical features is presented in Fig. 1, highlighting the phonon modes and hinting at a deep intrinsic relationship between ionic migration and EPC, both of which emerge from the same underlying soft and polar chemical bonding. Despite this connection, a unifying framework linking the lattice dynamics with important EPC and ionic transport properties across diverse MHP compositions remains notably absent. Here we introduce an orbital hybridization descriptor, $R$, to establish such a unified framework across representative classes of metal halide perovskites, targeting inorganic $CsPbI_3$, $CsSnI_3$, and $Cs_2AgBiBr_6$ as model systems spanning the Pb-based, Sn-based, and double perovskite family branches, respectively. By resolving the phonon-mode contributions to ionic transport and carrier scattering, we reveal that ion migration and EPC are governed by distinct phonon regimes but originate from a common electronic-structure framework via $R$, which quantitatively captures how chemical bonding simultaneously regulates lattice restoring forces and dielectric screening across diverse perovskite compositions (Fig. 1). These results establish orbital hybridization as a fundamentally important driver for coupled ionic and electronic transport in halide perovskites, providing a general framework for designing soft semiconductors with improved stability and transport properties.

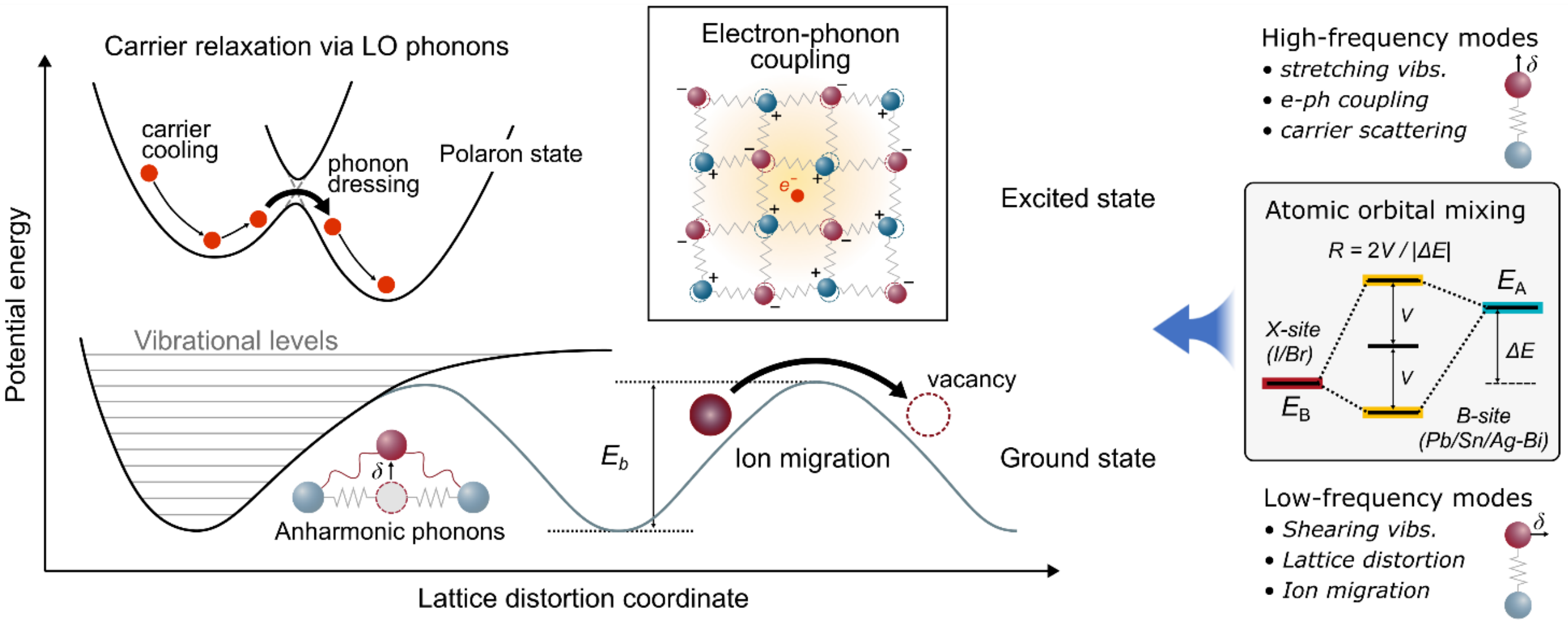

**Fig. 1 | Unified conceptual framework for electron–phonon coupling and ion migration.** The orbital-hybridization descriptor R=2V/｜ΔE｜ links chemical bonding to lattice dynamics. High-frequency LO stretching modes predominantly govern electron–phonon coupling and carrier relaxation by modulating orbital hybridization, whereas low-frequency shearing modes promote vacancy-mediated ion migration through lattice distortion. Although driven by distinct phonon modes, both phenomena arise from the same orbital-hybridization response of the metal–halide bonding network.

## Results and Discussion

We begin by outlining the physical models developed to study the relationship between anharmonic phonons, ion migration and electron–phonon coupling (Fig. 1). A dual-atom model(*32*) is implemented to define our orbital hybridization descriptor, *R*, that represents the tangent of twice the orbital mixing angle:

$$R = \frac{2V}{\Delta E} \tag{1}$$

where V is the interatomic orbital coupling (hopping) matrix element between the anion and cation orbitals, and ΔE denotes their onsite energy separation. This description originates from the two-level linear combination of the atomic orbitals model, where the bonding–antibonding splitting is determined by the competition between orbital coupling and energy mismatch, with further details outlined in the ESI. Physically, V quantifies the probability amplitude for electron transfer between neighboring atomic orbitals and can be obtained from the off-diagonal Hamiltonian matrix elements in the Wannier representation:

$$V = \langle \phi_A | H | \phi_B \rangle \tag{2}$$

where $\phi_A$ and $\phi_B$ are the corresponding atomic-like Wannier orbitals. The onsite energy difference,

$$\Delta E = |E_A - E_B| \tag{3}$$

subsequently describes the energetic mismatch between interacting orbitals and can be extracted from the diagonal elements of the Wannier Hamiltonian or orbital-projected electronic structure. Together, we propose that V and ΔE provide a quantitative measure of the tendency toward metal–halide orbital mixing and establish a direct link between electronic hybridization, lattice restoring forces, and phonon-mediated dynamic processes in the MHP lattice. The restoring forces and phonon-mediated dynamic processes directly govern the ion migration barrier, $E_b$, and the microscopic electron–phonon coupling strength represented by the matrix element, $g_{mnv}$, where $m$ and $n$ donate the electronic bands and $v$ is the phonon branches. Importantly, R is not intended as a direct descriptor of either migration barrier or EPC strength, but rather as an electronic-structure parameter that governs the orbital response to lattice perturbations. Furthermore, R provides predictive power for both short-range lattice dynamics and long-range Fröhlich interactions because local orbital hybridization ultimately determines the macroscopic dielectric response.

We next deploy this framework to describe crystalline MHPs and validate its accuracy, adopting the archetypal cubic phase of $CsPbI_3$, $CsSnI_3$ and $Cs_2AgBiBr_6$ (Fig. S2) for all calculations. Furthermore, at 300 K the systems are dynamically stabilized by anharmonic effects, which renormalized the phonon frequencies and eliminate imaginary modes, as discussed later. The fully-relaxed lattice constants are 6.37, 6.27 and 11.48 Å for $CsPbI_3$, $CsSnI_3$ and $Cs_2AgBiBr_6$, respectively, in agreement with previous studies(*34*–*36*). The calculated Perdew-Burke-Ernzerhof (PBE) electronic band structures are depicted in Fig. 2a, with both $CsPbI_3$ and $CsSnI_3$ being direct bandgap semiconductors, whereas $Cs_2AgBiBr_6$ exhibits an indirect bandgap, in good agreement with previous reports(*34–36*). The Wannier-interpolated bands reproduce the PBE band structures almost exactly, confirming the accuracy of the Wannier representation used in the electron–phonon calculations that follow. Further validation of the Wannier interpolation is provided in the Supplementary Information Fig. S3-S4. Hybrid HSE+SOC calculations were also performed to assess the influence of the exchange–correlation functional and, while HSE+SOC enlarges the band gaps, it leaves the band dispersions and orbital characters near the band edges essentially unchanged (Supplementary Fig. S5). This indicates that the bonding characteristics relevant to the present analysis are well captured by PBE and descriptor R is robust under the PBE functional. All EPC calculations were consistently performed using the PBE electronic structure within the EPW framework.

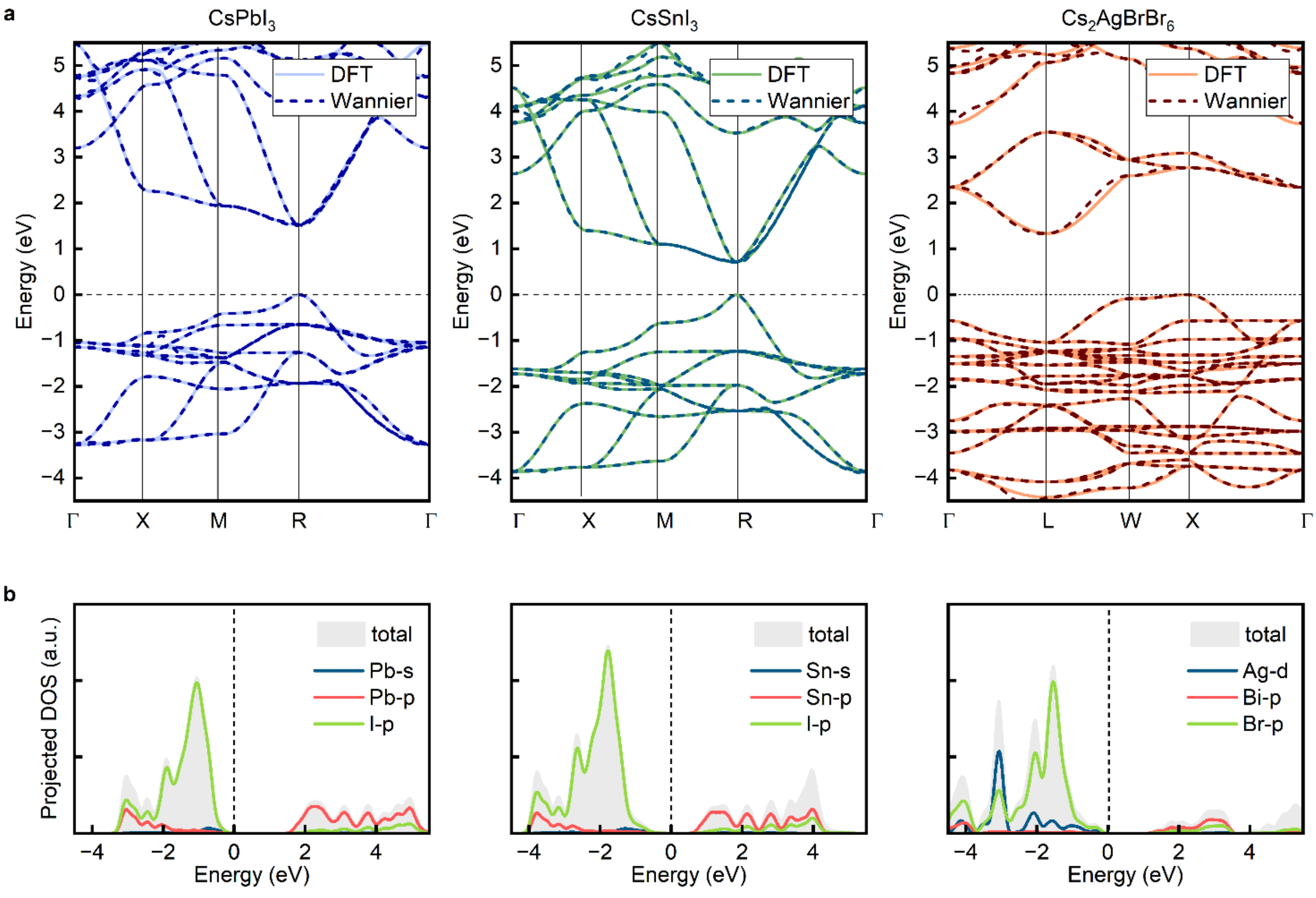

**Fig. 2 | Electronic structures of representative halide perovskites. a,** Electronic band structures of $CsPbI_3$, $CsSnI_3$ and $Cs_2AgBiBr_6$ calculated using density functional theory (solid lines) and Wannier interpolation (dashed lines). **b,** Corresponding projected densities of states

(DOS) for the three compounds.

To elucidate the orbital contributions near the band edges, we calculated the projected densities of states (PDOS), as shown in Fig. 2b. For the single perovskites, the valence-band maximum (VBM) is dominated by antibonding interactions between Pb-*s* and I-*p* orbitals, whereas the conduction-band minimum (CBM) is primarily derived from Pb-*p* states. In contrast, for the double perovskite $Cs_2AgBiBr_6$, the VBM mainly originates from antibonding Ag-*d* and Br-*p* states, while the CBM is dominated by Bi-*p* states, as shown in Fig. S3. These electronic structures are consistent with previous theoretical studies(*37*, *38*).

Calculating R we find that it exhibits a clear material dependence, with the largest value obtained for Bi-Br interactions in $Cs_2AgBiBr_6$ (R=1.69), followed by Sn-I (R=0.71) and Pb-I (R=0.48). This trend originates from the combination of enhanced V=1.01 eV and reduced ΔE=1.08 eV for Bi–Br interactions. Although Bi-*s*/Br-*s* orbitals are not directly responsible for the band-edge states, their strong hybridization reflects the intrinsic electronic coupling within the Bi–Br framework, which governs the local bonding environment and lattice response. The full quantitative details can be found in Tables S1 - S3.

Considering now the lattice dynamics, phonon dispersion calculations accounting for finite-temperature anharmonic effects (at 300 K and 400 K) are displayed in Fig. 3a, with their corresponding projected DOS shown in Fig. 3b. Notably, the imaginary frequencies present in the harmonic approximation (0 K) disappear in the finite temperature model, indicating the lattice is dynamically stabilized by strong anharmonicity in all three systems. The low-frequency shearing modes and acoustic phonon branches, as depicted in Fig. 3b, of $CsPbI_3$ and $Cs_2AgBiBr_6$ remain nearly unchanged between 300 K and 400 K, while $CsSnI_3$ exhibits pronounced phonon softening, i.e. decreasing frequencies, with rising temperature. These low-frequency phonons, including octahedral bending and tilting modes, possess intrinsically weak restoring forces. The displacement patterns of these modes are further shown below to exhibit a large projection onto the ion-migration coordinate, allowing lattice vibrations to efficiently drive the structural distortion required for ionic hopping. This directional correspondence provides a microscopic mechanism for why low-frequency phonons dominate ion migration. Conversely, high-frequency LO phonons – key for Fröhlich-type electron–phonon coupling – primarily originate from stretching vibrational modes that induce local polar fluctuations in the crystal.

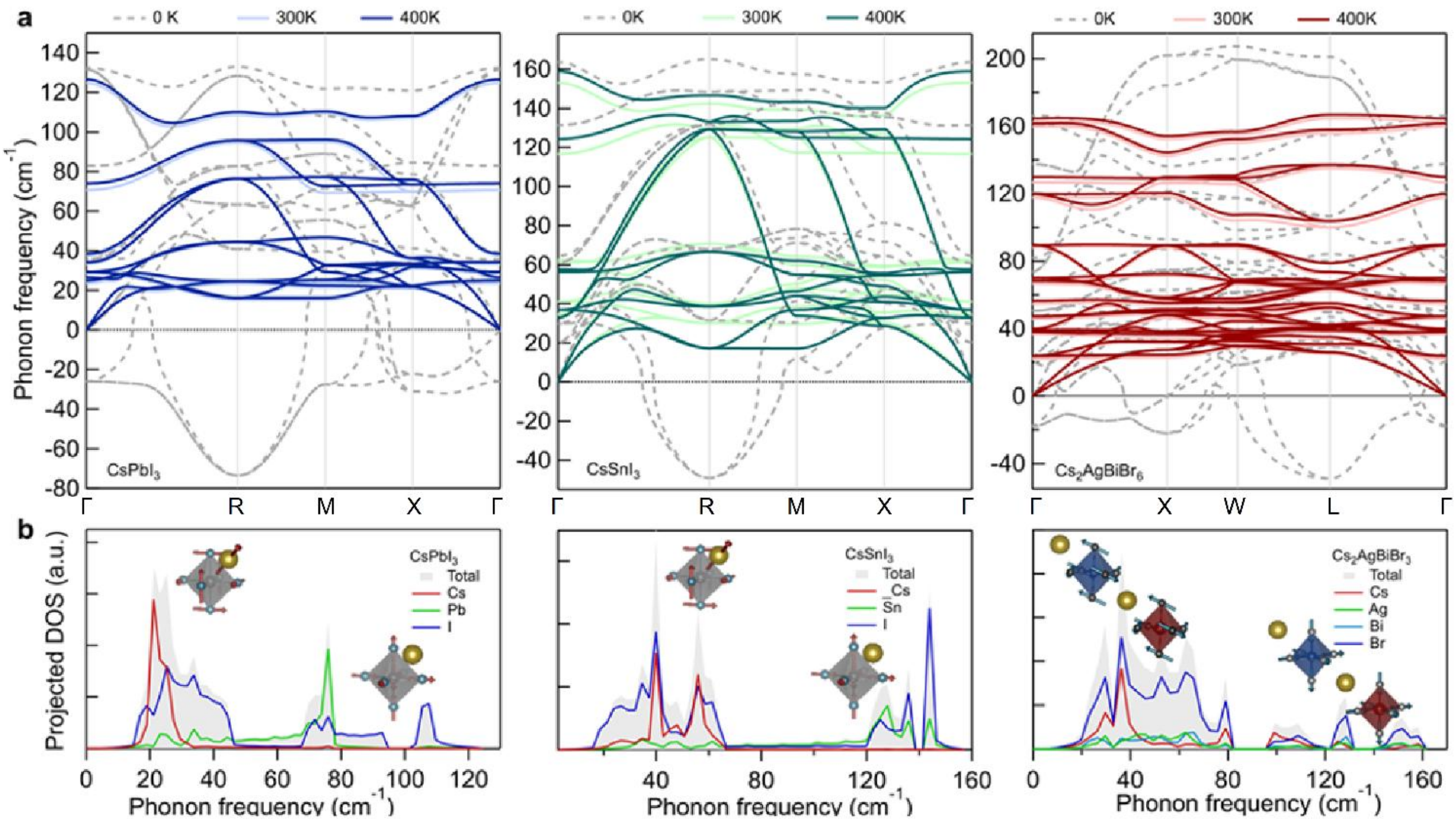

**Fig. 3 | Anharmonic lattice dynamics in representative halide perovskites. a,** Temperature-dependent phonon dispersions of $CsPbI_3$, $CsSnI_3$ and $Cs_2AgBiBr_6$ at 0 K (black dashed lines), 300 K (blue solid lines) and 400 K (red solid lines), highlighting the anharmonic renormalization of the phonon spectra. **b,** Corresponding phonon projected densities of states (PDOS) at 300 K.

On this basis, we can systematically establish the connections between *R*, ion migration and electron–phonon coupling by employing shearing and stretching vibrational modes, respectively. We thus consider how R responds to mode-specific lattice perturbations in the disparate compositions, through $\Delta R(\delta)=R(\delta)-R_0$. Here, the $R(\delta)$ and $R_0$ are respectively the displacement ($\delta$) and equilibrium state of R. For distortions involving bond stretching and compression, the average $\Delta R(\delta)$ associated with the Sn-I orbital pair (average $\Delta R(\delta)=0.09$) is smaller than that of the Pb-I pair (average $\Delta R(\delta)=0.14$), indicating that the Sn-I hybridization is less sensitive to longitudinal bond-length fluctuations. In contrast, for shear-like distortions, the Sn-based perovskite ($\Delta R(\delta)=0.10$) exhibits a substantially larger change than that of $CsPbI_3$ ($\Delta R(\delta)=0.03$), while the double perovskite (average $\Delta R(\delta)=1.11$ and $\Delta R(\delta)=0.98$ for stretching/compression and shearing, respectively) shows the largest variation among all three compositions. These mode-dependent responses suggest that the ionic migration barriers follow the sequence $Cs_2AgBiBr_6$ > $CsSnI_3$ > $CsPbI_3$, whereas the EPC strength follows $Cs_2AgBiBr_6$ > $CsPbI_3$ > $CsSnI_3$. Further details of the results can be found in Tables S4 – S12. The opposite ordering of the single perovskites therefore arises from the different sensitivity of the orbital hybridization to shear-like and stretching-like lattice distortions, which respectively couple most strongly to the phonon modes responsible for ion migration and Fröhlich-type EPC.

We verify these numerical determinations through a combination of climbing-image nudged elastic band (CI-NEB), electron linewidth calculations, and low-temperature photoluminescence spectroscopy. Previous experimental and theoretical studies have consistently demonstrated that halide anions are the primary mobile species in halide perovskites under ambient conditions(*13*, *20*, *22*, *33*). Accordingly, we restrict our analysis to the phonon modes with substantial halide-anion character in the projected phonon density of states. Through spectral analysis, we thus assess the individual and cumulative contributions of distinct phonons to ion migration (evaluated using projection analysis; see Methods) and visualize which specific vibrational modes drive or assist ionic diffusion. From Fig. 4a, we can see that the cumulative contribution of phonon modes from 0 to 40 $cm^{-1}$ approaches 100% for $CsPbI_3$, indicating that the ion migration coordinate is predominantly coupled to low frequency bending and tilting modes. These soft lattice vibrations can efficiently activate the structural distortions required for ionic displacement. In contrast, the corresponding contributions from modes below 40 $cm^{-1}$ are only about 80% and 30% for $CsSnI_3$ and $Cs_2AgBiBr_6$, respectively, suggesting that ion migration in these systems requires participation for higher frequency lattice distortions and is therefore associated with stronger structural constraints. Combined with the bonding strength descriptor R, these phonon projection analyses reveal an increasing difficulty for ion migration from $CsPbI_3$ to $CsSnI_3$ to $Cs_2AgBiBr_6$.

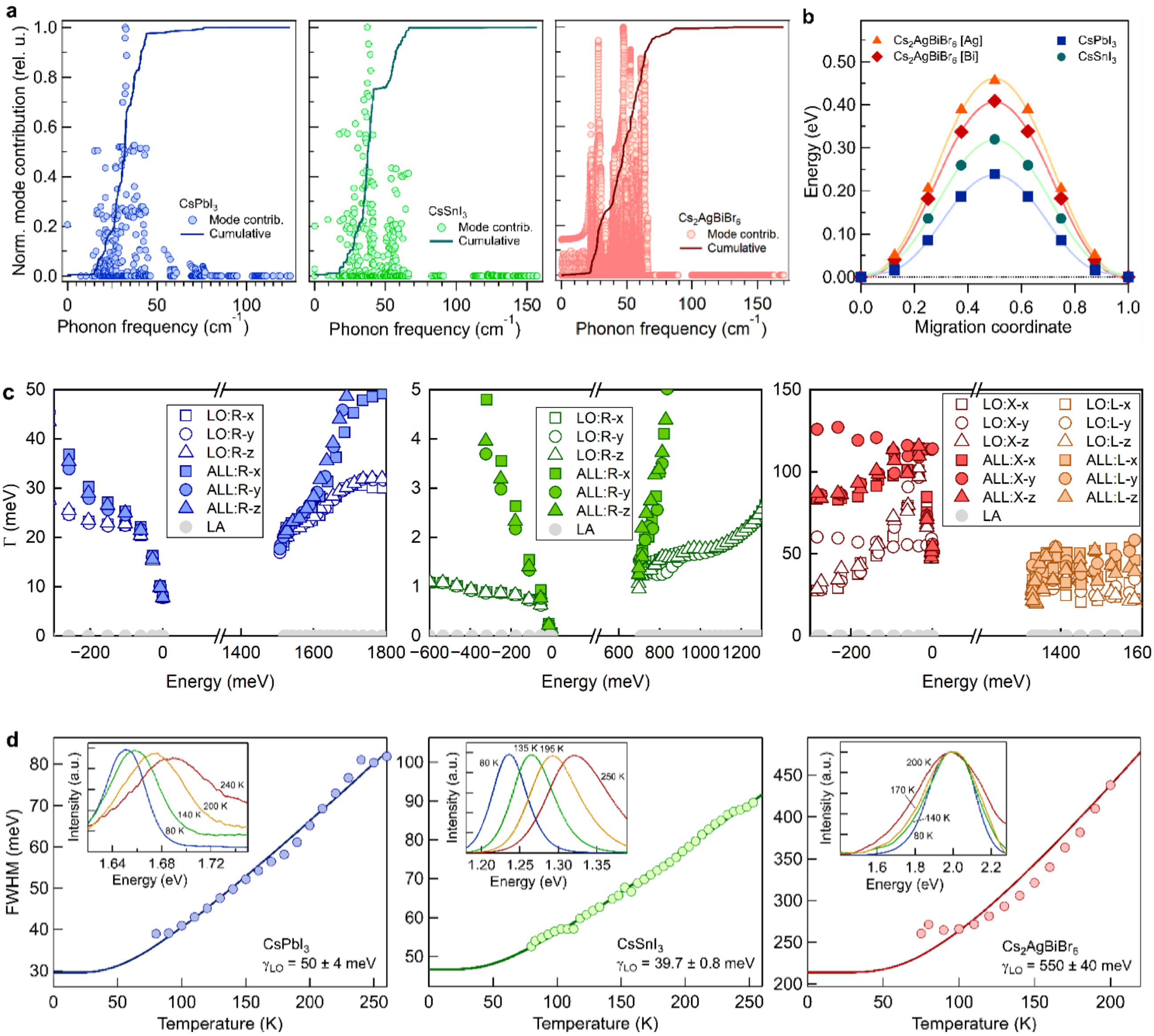

**Fig. 4 | Phonon-mode-resolved ion migration and electron–phonon coupling in halide perovskites. a,** Scatter plots and cumulative contribution plots of halogen migration contributions from different phonon modes (including all q-space phonons) in $CsPbI_3$, $CsSnI_3$, and $Cs_2AgBiBr_6$. **b,** Calculated migration barriers of different halogen atoms in these compositions. **c,** Electron–phonon interaction induced energy-dependent electronic linewidths ($\Gamma_{e\text{-}ph}$) at the lowest conduction band (LCB) and the highest valence band (HVB). The energy is referenced to the matching band edge. Filled symbols represent the total linewidth, open symbols denote the polar longitudinal optical (LO) phonon contribution to the electronic linewidths, different shapes correspond to different Cartesian direction, and the grey circles indicate the contribution from longitudinal acoustic (LA) phonon modes. **d,** Experimentally determined steady-state photoluminescence emission full widths at half maximum (FWHM) as a function of temperature, with an example of measured normalized data spectra provided in the insets for several different temperatures. The solid line is a fit made with **eq 4**. The determined contributions to the line width broadening induced by LO phonon (Fröhlich interactions) coupling are provided in the respective insets.

To quantitatively evaluate this trend, we further calculated the ion migration barriers using the CI-NEB methods, as shown in Fig. 4b. In excellent agreement with the predictions made via the changes in R(δ) and a cumulative phonon contribution approach, the numerical barriers are 0.239 eV, 0.320 eV and 0.408 eV for $CsPbI_3$, $CsSnI_3$ and $Cs_2AgBiBr_6$, respectively. Experimentally, isolating the intrinsic contribution to ion migration barriers remains challenging because numerous extrinsic factors, including crystal quality and grain boundaries, can influence ion migration. Moreover, the three compositions exhibit distinct levels of synthetic complexity, making it difficult to achieve comparable crystal quality across all samples. Therefore, experimental validation under strictly controlled conditions would be highly desirable to further corroborate our theoretical findings.

Based on the above analyses, a unified link between phonon occupation, orbital hybridization, and ion migration can be established. Given that low-frequency phonon modes dominate halide-ion migration, the hopping probability increases with the occupation of these modes, leading to a thermally activated enhancement of ion migration with increasing temperature. This behavior can be consistently rationalized using the descriptor R. As ΔE decreases and V increases, stronger orbital hybridization enhances both the static bonding strength and the sensitivity of orbital interactions to lattice perturbations, resulting in larger R and dynamic variation ΔR(δ). While the increased static hybridization strengthens metal–halide bonding and raises the migration barrier, the enhanced ΔR(δ) indicates a stronger coupling between low-frequency shearing vibrations and the migration coordinate, allowing phonons to more effectively modulate ionic motion.

To capture the overall impact of electron–phonon interactions in carrier scattering, we compute the electron-phonon self-energy, which quantifies the broadening of electronic states induced by phonon-mediated scattering. Fig. 4c presents the electron linewidth distributions within a narrow energy range around the band edges. For the direct-bandgap perovskites $CsPbI_3$ and $CsSnI_3$, the linewidths are nearly isotropic along different crystallographic directions, whereas $Cs_2AgBiBr_6$ exhibits larger, more anisotropic self-energies near both the lowest conduction band (LCB: L point) and the highest valence band (HVB: X point), indicating stronger carrier scattering and shorter lifetimes. This enhanced scattering originates from stronger long-range polar coupling associated with LO phonons, consistent with the larger R and dynamical R(δ) in $Cs_2AgBiBr_6$. By comparison, the stronger covalent character of the Sn–I bond in $CsSnI_3$ weakens the long-range polar nature of lattice vibrations and suppresses Fröhlich coupling, resulting in reduced carrier scattering. In fact, comparing the full electron linewidth and the linewidth arising solely from the long-range Fröhlich interaction demonstrates that carrier scattering is overwhelmingly dominated by the LO-Fröhlich channel. On the other hand, the negligible contribution from LA phonons confirms that short-range deformation-potential scattering is not the dominant carrier-scattering mechanism near the band edges. The material dependence of the Fröhlich interaction strength follows the same electronic-structure trend identified above: $Cs_2AgBiBr_6$ > $CsPbI_3$ > $CsSnI_3$. Enhanced R and R(δ) increase lattice polarizability, strengthening the macroscopic polarization field associated with LO vibrations.

We experimentally validate these theoretical predictions by infering the EPC strengths in solution-processed $CsPbI_3$, $CsSnI_3$, and $Cs_2AgBiBr_6$ thin films (see Methods) from temperature-dependent PL linewidths (Fig. 4d) using a phonon-mediated broadening model.(*15*, *39*):

$$\Gamma(T) = \Gamma_0 + \gamma_{AC}T + \frac{\gamma_{LO}}{e^{\frac{\hbar\omega_{LO}}{k_B T}} - 1} \tag{4}$$

where $\gamma_{LO}$ and $\gamma_{AC}$ represent LO- and acoustic-phonon coupling strengths, respectively, and $\Gamma_0$ is the inherent linewidth of the studied material at 0 K. Given the LO contribution dominates the linewidth broadening near the band edges, with only the negligible contribution of acoustic modes, the second term can be omitted to estimate the upper limit of Fröhlich coupling strength here. Using LO phonon energies obtained from Raman measurements (Fig. S12), we determine $\gamma_{LO}$ values of 50 ± 4 meV, 39.7 ± 0.8 meV, and 550 ± 50 meV for $CsPbI_3$, $CsSnI_3$, and $Cs_2AgBiBr_6$, respectively. The experimentally extracted trend, $Cs_2AgBiBr_6$ > $CsPbI_3$ > $CsSnI_3$, agrees with our calculations and follows the same evolution of the orbital-hybridization descriptor. Specifically, larger static R and ΔR(δ) responses enhance the sensitivity of electronic states to LO stretching vibrations, leading to stronger modulation of carrier states and enhanced EPC.

## Conclusion

Our work establishes a unified microscopic framework connecting lattice anharmonicity, electron–phonon coupling, and ion migration across representative metal halide perovskites. Although ionic transport and carrier scattering are governed by distinct regions of the phonon spectrum – low-frequency bending and tilting modes for ion migration and high-frequency longitudinal optical phonons for Fröhlich coupling – we trace their origin to the same underlying electronic structure. Specifically, orbital hybridization simultaneously determines the lattice restoring forces, dielectric screening, and bond polarity, thereby coupling anharmonic lattice dynamics to both ionic and electronic transport. By developing the descriptor $R=2V/\Delta E$, we demonstrate that orbital hybridization provides a common basis for linking chemical bonding, lattice dynamics, carrier scattering, and ion migration across Pb-based, Sn-based, and double-perovskite systems. This framework explains why enhanced orbital hybridization simultaneously promotes ionic migration barriers, and modifies long-range polar electron–phonon interactions through changes in dielectric screening. The excellent agreement between first-principles calculations and temperature-dependent photoluminescence measurements further supports the generality of this mechanism. More broadly, our work shifts the perspective from treating electron–phonon coupling and ion migration as independent phenomena to viewing them as complementary manifestations of the same electronic-structure. This unified understanding provides a rational strategy for simultaneously optimizing charge transport and operational stability through orbital engineering, offering general design principles for next-generation halide perovskites and other soft polar semiconductors.

## Supporting Information

Supporting Information is available from the XXXX or from the author.

## Acknowledgements

The authors also acknowledge the financial support from the National Natural Science Foundation of China (Grant Nos. 62404107, 62304111, 62474097, 22579136, 225B2914), the S&T Program of Energy Shaanxi Laboratory (ESLB202438), Youth project in natural science and engineering technology (2023SYJ15), Xi'an Jiaotong University Youth Innovation Team (xtr052025016, xtr072024024), and the China Fundamental Research Funds for the Central Universities, the Project of the State Key Laboratory of Flexible Electronics (Nos. GDX2022010009, GZR2023010041), S. Y. C. acknowledges support from EPSRC [EP/V062654/1]. J.A.S. and C. V. acknowledges financial support from the Australian Research Council (DE230100173, DP260103534, DE220101147). This research was supported by the UQ Amplify Fellowship scheme from The University of Queensland.

**Conflict of Interest**

The authors declare no conflict of interest.

**References**

1. T. M. Brenner, D. A. Egger, L. Kronik, G. Hodes, D. Cahen, Hybrid organic—inorganic perovskites: low-cost semiconductors with intriguing charge-transport properties. *Nat. Rev. Mater.* **1**, 15007 (2016).

2. Y. Fu, H. Zhu, J. Chen, M. P. Hautzinger, X.-Y. Zhu, S. Jin, Metal halide perovskite nanostructures for optoelectronic applications and the study of physical properties. *Nat. Rev. Mater.* **4**, 169–188 (2019).

3. S. D. Stranks, H. J. Snaith, Metal-halide perovskites for photovoltaic and light-emitting devices. *Nat. Nanotechnol.* **10**, 391–402 (2015).

4. A. J. Ramadan, R. D. J. Oliver, M. B. Johnston, H. J. Snaith, Methylammonium-free wide-bandgap metal halide perovskites for tandem photovoltaics. *Nat. Rev. Mater.* **8**, 822–838 (2023).

5. J. Huang, Y. Yuan, Y. Shao, Y. Yan, Understanding the physical properties of hybrid perovskites for photovoltaic applications. *Nat. Rev. Mater.* **2**, 17042 (2017).

6. B. A. Rosales, K. Schutt, J. J. Berry, L. M. Wheeler, Leveraging Low-Energy Structural Thermodynamics in Halide Perovskites. *ACS Energy Lett.* **8**, 1705–1715 (2023).

7. N. Xu, X. Qi, Z. Shen, L. Hu, J. Lv, Y. Zhong, B. Wang, Z. Zou, Point defects in metal halide perovskites. *Nat. Rev. Phys.* **7**, 554–564 (2025).

8. C. Gehrmann, D. A. Egger, Dynamic shortening of disorder potentials in anharmonic halide perovskites. *Nat. Commun.* **10**, 3141 (2019).

9. J. A. Steele, Atomistic origins of anharmonic lattice dynamics and thermal expansion in perovskite photovoltaics. *Nat. Energy* **11**, 372–386 (2026).

10. S. Caicedo-Dávila, A. Cohen, S. G. Motti, M. Isobe, K. M. McCall, M. Grumet, M. V. Kovalenko, O. Yaffe, L. M. Herz, D. H. Fabini, D. A. Egger, Disentangling the effects of structure and lone-pair electrons in the lattice dynamics of halide perovskites. *Nat. Commun.* **15**, 4184 (2024).

11. A. Singh, D. B. Mitzi, Emergence of melt and glass states of halide perovskite semiconductors. *Nat. Rev. Mater.* **10**, 211–227 (2025).

12. J. M. Frost, Calculating polaron mobility in halide perovskites. *Phys Rev B* **96**, 195202 (2017).

13. C. Eames, J. M. Frost, P. R. F. Barnes, B. C. O'Regan, A. Walsh, M. S. Islam, Ionic transport in hybrid lead iodide perovskite solar cells. *Nat. Commun.* **6**, 7497 (2015).

14. J. M. Azpiroz, E. Mosconi, J. Bisquert, F. De Angelis, Defect migration in methylammonium lead iodide and its role in perovskite solar cell operation. *Energy Environ. Sci.* **8**, 2118–2127 (2015).

15. A. D. Wright, C. Verdi, R. L. Milot, G. E. Eperon, M. A. Pérez-Osorio, H. J. Snaith, F. Giustino, M. B. Johnston, L. M. Herz, Electron–phonon coupling in hybrid lead halide perovskites. *Nat. Commun.* **7**, 11755 (2016).

16. L. M. Herz, Charge-Carrier Dynamics in Organic-Inorganic Metal Halide Perovskites, *Annual Review of Physical Chemistry*. **67** (2016)pp. 65–89.

17. B. Monserrat, N. D. Drummond, R. J. Needs, Anharmonic vibrational properties in periodic systems: energy, electron-phonon coupling, and stress. *Phys Rev B* **87**, 144302 (2013).

18. H. Zhu, K. Miyata, Y. Fu, J. Wang, P. P. Joshi, D. Niesner, K. W. Williams, S. Jin, X.-Y. Zhu, Screening in crystalline liquids protects energetic carriers in hybrid perovskites. *Science* **353**, 1409–1413 (2016).

19. K. Miyata, D. Meggiolaro, M. T. Trinh, P. P. Joshi, E. Mosconi, S. C. Jones, F. De Angelis, X.-Y. Zhu, Large polarons in lead halide perovskites. *Sci. Adv.* **3**, e1701217.

20. Y. Yuan, J. Huang, Ion Migration in Organometal Trihalide Perovskite and Its Impact on

Photovoltaic Efficiency and Stability. *Acc. Chem. Res.* **49**, 286–293 (2016).

21. J. Thiesbrummel, J. V. Milić, C. Deibel, E. C. Garnett, S. Tao, T. Kirchartz, A. Guerrero, P. Cameron, W. Tress, M. Saiful Islam, B. Ehrler, Ion migration in perovskite solar cells. *Nat. Rev. Chem.* **10**, 179–195 (2026).

22. E. T. Hoke, D. J. Slotcavage, E. R. Dohner, A. R. Bowring, H. I. Karunadasa, M. D. McGehee, Reversible photo-induced trap formation in mixed-halide hybrid perovskites for photovoltaics. *Chem. Sci.* **6**, 613–617 (2015).

23. D. T. Limmer, N. S. Ginsberg, Photoinduced phase separation in the lead halides is a polaronic effect. *J. Chem. Phys.* **152**, 230901 (2020).

24. S. G. Motti, J. B. Patel, R. D. J. Oliver, H. J. Snaith, M. B. Johnston, L. M. Herz, Phase segregation in mixed-halide perovskites affects charge-carrier dynamics while preserving mobility. *Nat. Commun.* **12**, 6955 (2021).

25. S. K. Gautam, M. Kim, D. R. Miquita, J. Bourée, B. Geffroy, O. Plantevin, Reversible Photoinduced Phase Segregation and Origin of Long Carrier Lifetime in Mixed-Halide Perovskite Films. *Adv. Funct. Mater.* **30**, 2002622 (2020).

26. L. A. Castriotta, M. Wang, X. Shi, B. D. Dou, L. T. Schelhas, J. Huang, Challenges, technological pathways and trade-offs of perovskite solar modules for long-term operation. *Nat. Energy* **11**, 534–546 (2026).

27. R. Claes, S. Poncé, G.-M. Rignanese, G. Hautier, Phonon-limited electronic transport through first principles. *Nat. Rev. Phys.* **7**, 73–90 (2025).

28. I. Mosquera-Lois, Y.-T. Huang, H. Lohan, J. Ye, A. Walsh, R. L. Z. Hoye, Multifaceted nature of defect tolerance in halide perovskites and emerging semiconductors. *Nat. Rev. Chem.* **9**, 287–304 (2025).

29. O. Yaffe, Y. Guo, L. Z. Tan, D. A. Egger, T. Hull, C. C. Stoumpos, F. Zheng, T. F. Heinz, L. Kronik, M. G. Kanatzidis, J. S. Owen, A. M. Rappe, M. A. Pimenta, L. E. Brus, Local Polar Fluctuations in Lead Halide Perovskite Crystals. *Phys. Rev. Lett.* **118**, 136001 (2017).

30. A. N. Beecher, O. E. Semonin, J. M. Skelton, J. M. Frost, M. W. Terban, H. Zhai, A. Alatas, J. S. Owen, A. Walsh, S. J. L. Billinge, Direct Observation of Dynamic Symmetry Breaking above Room Temperature in Methylammonium Lead Iodide Perovskite. *ACS Energy Lett.* **1**, 880–887 (2016).

31. B. Yang, X. Wei, B. Cai, Y. Yang, X. Zhu, Y. Liu, J. Xia, L. Liu, K. Cao, W. Shen, P. Xia, S. Chen, S. Chen, J. Zhao, Machine Learning Accelerated Non-Adiabatic Molecular Dynamics Elucidates

Local Polarization Effects on Non-radiative Recombination in Halide Perovskites. *Adv. Sci.*, e75903 (2026).

32. B. Cai, Y. Ma, B. Yang, Y. Liu, J. Xia, X. Chen, Z. Li, M. Ju, A New Descriptor for Complicated Effects of Electronic Density of States on Ion Migration. *Adv. Funct. Mater.* **33**, 2300445 (2023).

33. A. N. Arber, Vikram, F. C. Mocanu, M. S. Islam, Ion Migration and Dopant Effects in the Gamma-$CsPbI_3$ Perovskite Photovoltaic Material: Atomistic Insights through *Ab Initio* and Machine Learning Methods. *Chem. Mater.* **37**, 4416–4424 (2025).

34. K.-C. Zhang, C. Shen, H.-B. Zhang, Y.-F. Li, Y. Liu, Effect of quartic anharmonicity on the carrier transport of cubic halide perovskites CsSnI 3 and CsPbI 3. *Phys. Rev. B* **106**, 235202 (2022).

35. M. Zafar, M. Muddassir, A. Ali, M. Shakil, I. H. El Azab, Comparative analysis of band gap using different approximations, structural, mechanical and optical behaviour analysis of lead free double halide perovskites Cs 2 AgBiBr 6 using DFT approach. *Solid State Commun.* **397**, 115825 (2025).

36. M. A. Oufakir, R. T. Alqahtani, Y. Chrafih, A. Ajbar, Next-gen solar: revealing the promise of CsPbI3/CsSnI3 tandem cells. *Sol. Energy* **298**, 113665 (2025).

37. G. Volonakis, M. R. Filip, A. A. Haghighirad, N. Sakai, B. Wenger, H. J. Snaith, F. Giustino, Lead-Free Halide Double Perovskites via Heterovalent Substitution of Noble Metals. *J. Phys. Chem. Lett.* **7**, 1254–1259 (2016).

38. A. Walsh, D. O. Scanlon, S. Chen, X. G. Gong, S. Wei, Self-Regulation Mechanism for Charged Point Defects in Hybrid Halide Perovskites. *Angew. Chem. Int. Ed.* **54**, 1791–1794 (2015).

39. J. A. Steele, P. Puech, M. Keshavarz, R. Yang, S. Banerjee, E. Debroye, C. W. Kim, H. Yuan, N. H. Heo, J. Vanacken, A. Walsh, J. Hofkens, M. B. J. Roeffaers, Giant Electron–Phonon Coupling and Deep Conduction Band Resonance in Metal Halide Double Perovskite. *ACS Nano* **12**, 8081–8090 (2018).

# Supplementary Information: A Unified Description of Electron-Phonon Coupling and Ion Migration in Metal Halide Perovskites

Bo Cai[1,+,*], Yan Yang[1,+], Yoshiki Sugai[2], Maddison Wiles[3], Dongxu He[4], Yang Yang[1], Junmin Xia[1,*], Shufen Chen[1,*], Carla Verdi[3,*], Siyu Chen[5,6,7,*], Nan Zhang[8], Ming-Gang Ju[9], Chao Liang[8,*], and Julian A. Steele[2,3,*]

*1 State Key Laboratory of Flexible Electronics (LoFE) & Institute of Advanced Materials (IAM), Nanjing University of Posts and Telecommunications, Nanjing, 210023, P. R. China.*

*2 Australian Institute for Bioengineering and Nanotechnology, The University of Queensland, Brisbane, QLD, 4072 Australia.*

*3 School of Mathematics and Physics, The University of Queensland, Brisbane, QLD, 4072 Australia.*

*4 School of Chemical Engineering, The University of Queensland, Brisbane, QLD, 4072 Australia.*

*5 TCM Group, Cavendish Laboratory, University of Cambridge, Cambridge CB3 0HE, UK*

*6 Department of Materials Science and Metallurgy, University of Cambridge, Cambridge CB3 0FS, UK*

*7 European Theoretical Spectroscopy Facility, Institute of Condensed Matter and Nanosciences, Université catholique de Louvain, Louvain-laNeuve, 1348, Belgium*

*8 State Key Laboratory of Electrical Insulation and Power Equipment, MOE Key Laboratory for Nonequilibrium Synthesis and Modulation of Condensed Matter, School of Physics, Xi'an Jiaotong University, Xi'an, 710049, P. R. China*

*9 Key Laboratory of Quantum Materials and Devices of Ministry of Education, School of Physics, Southeast University, 211189 Nanjing, China.*

Email: iambcai@njupt.edu.cn; iamjmxia@njupt.edu.cn; iamsfchen@njupt.edu.cn; c.verdi@uq.edu.au; sc2090@cam.ac.uk; chaoliang@xjtu.edu.cn; julian.steele@uq.edu.au.

**Table of Contents**

## Methods

## Computational Details

### DFT calculations

All first-principles calculations were performed using the Quantum ESPRESSO package(*1*, *2*) within the framework of density functional theory (DFT) using a plane-wave basis set(*3*) and norm-conserving pseudopotentials(*4*) to describe the electron–ion interactions. Convergence tests were carried out with respect to the plane-wave kinetic energy cutoff, and the kinetic energy and charge density cutoffs were set to 70 Ry and 280 Ry, respectively. Brillouin-zone integrations were performed using Monkhorst–Pack k-point meshes(*5*) of 2 × 2 × 2 for supercell. Supercells of 2 × 2 × 2 were employed for $CsPbI_3$, $CsSnI_3$, and $Cs_2AgBiBr_6$. Structural relaxations were performed using the Perdew–Burke–Ernzerhof (PBE) exchange–correlation functional(*6*), while electronic band structures were calculated using both PBE (without spin-orbital coupling) the Heyd–Scuseria–Ernzerhof (HSE) hybrid functional(*7*) including spin–orbit coupling (SOC). The convergence criteria for the total energy and atomic forces were set to $10^{-5}$ eV and 0.02 eV $Å^{-1}$, respectively. All electron-phonon coupling calculations and descriptor analyses are based on PBE functional.

### Wannier interpolation

To efficiently obtain electronic band structures, densities of states, and electron–phonon matrix elements on dense k-point grids, maximally localized Wannier functions (MLWFs) were constructed using the Wannier90 package(*8*). Within the Wannier representation, Bloch wavefunctions within a selected energy window are transformed into a set of localized Wannier orbitals through a unitary transformation in reciprocal space. This procedure maps the Bloch Hamiltonian electronic structure onto an effective real-space Hamiltonian, $H_{mn}(R)$, which decays rapidly with the lattice vector $\boldsymbol{R}$.

The Wannier representation enables efficient interpolation of the electronic band structure and density of states on dense k-point grids. For $CsPbI_3$ and $CsSnI_3$, the Wannier basis consisted primarily of Pb (or Sn) *s*, *p,* and *d* orbitals together with I *s* and *p* orbitals, whereas for $Cs_2AgBiBr_6$, Ag *s*, *p*, *d* orbitals, Bi *s*, *p*, *d* orbitals and Br *s* and *p* orbitals were included. Stable MLWFs were obtained by optimizing the disentanglement energy windows and convergence parameters. The quality of the Wannierization was assessed by comparing the Wannier-interpolated band structures with the corresponding first-principles bands in the vicinity of the Fermi level. In addition, the spatial spreads of the MLWFs and the decay behavior of the real-space Hamiltonian, $H(\boldsymbol{R})$, were examined to verify the localization of the Wannier functions and the reliability of the interpolation, thereby ensuring the accuracy of the subsequent electron–phonon coupling and transport calculations.

**Phonon calculations**

Harmonic phonon calculations were performed using density functional perturbation theory (DFPT) as implemented in the Quantum ESPRESSO package(*1*, *2*). Within the DFPT framework, the dynamical matrices were obtained by solving the linear response of the electronic density to infinitesimal atomic displacements, enabling direct calculations of phonon frequencies and vibrational eigenmodes. After the ground-state electronic structure had been converged, the dynamical matrices were computed using the ph.x module, and the phonon dispersion relations were subsequently generated with matdyn.x. For $CsPbI_3$, $CsSnI_3$, and $Cs_2AgBiBr_6$, phonon frequencies at the Γ point were first evaluated to examine the dynamical stability of the long-wavelength limit. The acoustic sum rule (ASR) was enforced to eliminate spurious frequencies in the acoustic branches, and non-analytic corrections (NACs) were included for these polar materials to account for the longitudinal optical–transverse optical (LO–TO) splitting.

The ZG. x code implementing the anharmonicity-special displacement method (A-SDM) procedure with the frozen-phonon method is available at the EPW/ZG module(*9*). A-SDM phonon dispersions were obtained using 2x2x2 supercells for the perovskite structures. At each iteration, including iteration 0 for the polymorphous structure, we enforced the crystal's symmetry operations on the IFCs. Setting the mixing parameter B to 0.5, we found that a couple of iterations is enough to obtain reasonable convergence and a maximum of 3-4 iterations to achieve full convergence. We employed 10 iterations as default. Long-range corrections in A-SDM phonon dispersions and ZG displacements were accounted for using the Born effective charges and dielectric constants of the polymorphous structures.

**Phonon contribution to ion migration**

To quantify the coupling between lattice vibrations and the ion-migration pathway, the phonon eigenvectors were projected onto the migration coordinate obtained from the corresponding migration pathway. For each phonon mode $v$, the individual contribution was defined as the normalized projection:

$$I_v = \frac{\left|e_v \cdot d_{mig}\right|^2}{\max(\left|e_v \cdot d_{mig}\right|^2)}$$

where $e_v$ is the normalized phonon eigenvector and $d_{mig}$ is the normalized displacement vector along the ion-migration coordinate. $I_v$ represents the fractional contribution of an individual phonon mode to the migration coordinate.

The cumulative phonon contribution is then calculated as:

$$C(\omega_c) = \frac{\sum_{\omega_v < \omega_c} \left|e_v \cdot d_{mig}\right|^2}{\sum_v \left|e_v \cdot d_{mig}\right|^2}$$

where $\omega_v$ is the frequency of phonon mode $v$. $C(\omega_c)$ describes the total contribution of phonon modes within a specific frequency range to the migration coordinate.

**Migration barrier calculations**

The climbing-image nudged elastic band (CI-NEB) method(*10*) was employed to determine the minimum energy pathways (MEPs) and the corresponding migration barriers for ionic diffusion. The initial and final configurations of the vacancy migration process were first fully relaxed and used as the two endpoints of the migration pathway. Intermediate images were then generated by linear interpolation between the two endpoints. For $CsPbI_3$, $CsSnI_3$, and $Cs_2AgBiBr_6$, each migration pathway was constructed using five intermediate images. During the NEB optimization, the total force acting on each image was decomposed into components parallel and perpendicular to the reaction pathway. The true potential force was retained only in the direction perpendicular to the path, while artificial spring forces were applied along the tangent direction to maintain an approximately uniform spacing between adjacent images, allowing the pathway to converge toward the MEP. To accurately locate the transition state, the climbing-image scheme was applied to the highest-energy image by removing the spring force along the reaction coordinate and reversing the tangential component of the potential force, driving the image toward the first-order saddle point. All CI-NEB calculations were performed with the lattice parameters fixed, allowing only the atomic positions to relax. The convergence criterion for the maximum residual force on each image was set to 0.03 eV $Å^{-1}$, ensuring reliable determination of the migration barriers.

**COHP calculations**

To characterize the chemical bonding interactions in halide perovskites, crystal orbital Hamilton population (COHP) analysis(*11*, *12*) was performed to quantify the bonding characteristics between specific atomic pairs. COHP partitions the Hamiltonian matrix elements into contributions from selected orbital pairs, thereby resolving bonding and antibonding interactions as a function of energy. In the conventional COHP representation, negative and positive values correspond to bonding and antibonding interactions, respectively. Throughout this work, the negative COHP (−COHP) convention is adopted, such that positive values represent bonding contributions and negative values represent antibonding contributions. To further quantify the relative bond strengths, the integrated crystal orbital Hamilton population (ICOHP) was calculated, with the absolute value of ICOHP serving as a measure of bond strength. The converged ground-state electronic structures obtained from first-principles calculations were post-processed using the LOBSTER code(*13*), which projects the plane-wave wavefunctions onto a localized atomic-orbital basis. By comparing the −COHP spectra and ICOHP values of the key atomic pairs among different perovskite compositions, the evolution of bonding and antibonding interactions and their correlation with structural stability and ion migration barriers were systematically analyzed.

**Electron linewidth calculations**

Electron linewidths arising from electron–phonon interactions were calculated using the EPW code based on the Wannier-interpolated electronic structure, anharmonic phonon dispersions and electron-phonon matrix elements(*14*–*16*). After obtaining MLWFs, the electronic band energies $\varepsilon_{nk}$, phonon frequencies $\omega_{qv}$, and electron-phonon matrix elements $g_{mnv}(k, q)$ were interpolated from coarse electron $\boldsymbol{k}$ grid and phonon $\boldsymbol{q}$ grid to dense Brillouin-zone grids. The electron linewidth, corresponding to the imaginary part of the electron self-energy within the Migdal approximation, was evaluated by integrating the electron-phonon scattering contributions over the dense momentum grids as

$$\Gamma_{nk}(T) = 2 Im \Sigma_{nk}(T)$$

where $\Sigma_{nk}$ is the electron self-energy arising from electron–phonon scattering, $n$ and $k$ denote the electronic band index and wave vector, respectively, and $T$ is the temperature. The linewidth includes contributions from both phonon absorption and emission processes over all phonon branches and wave vectors. Owing to its direct relationship with the electron–phonon scattering rate, the electron linewidth provides a quantitative measure of the overall strength of electron–phonon interactions and was used throughout this work to characterize carrier scattering.

## Experimental Details

### Sample preparation

### Chemical

All materials, unless otherwise stated, were purchased from Sigma-Aldrich and were used as received. $PbI_2$ was purchased from TSI.

### $CsPbI_3$

The $CsPbI_3$ thin films were prepared from a precursor solution of CsI, $PbI_2$, and DMAI at a 1 : 1 : 1.2 equivalence ratio and dissolved in DMSO by stirring. Before use, the solution was filtered with polytetrafluoroethylene (PTFE) syringe filters (0.22 µm). 35µl of the product was dropped onto a glass substrate (1.25 $cm^2$) and spin-coated at 4000 rpm for 60s with an acceleration rate of 2000 rpm. The glass substrates (1.25 $cm^2$) were cleaned sequentially by sonication in isopropanol, acetone, and isopropanol for 20 min each. The substrate surface was further treated with UV-Ozone for 15 min before use. To remove any remaining solvents, the yellow δ-$CsPbI_3$ phase film was then annealed at 120 °C for 2 min. To induce a phase transformation from the yellow δ-$CsPbI_3$ phase to the black γ-$CsPbI_3$ phase, the film was heated to 200°C under a nitrogen atmosphere and annealed for approximately 3 minutes. Due to the instability of the perovskite phase at room temperature and its sensitivity to humidity and oxygen, the yellow δ- $CsPbI_3$ phase film was annealed directly before cooling to the initial trial temperature of 80 K, thereby quenching the film and pinning the phase.

### $CsSnI_3$

For the precursor solution, iodine (0.665 g, Sigma-Aldrich) was fully dissolved in DMSO (0.75 ml, Sigma-Aldrich) first, then 3ml DMF (Sigma-Aldrich) was added to obtain a 0.7 M iodine solution. An excess of Sn powder (Sigma-Aldrich) (>1 g) was added to the solution, which was then stirred overnight in a nitrogen-filled glovebox to prepare a $SnI_2$ solution in DMF:DMSO (7:3 v/v). Then, 0.181 g of CsI (Sigma-Aldrich) was added to 1 ml of $SnI_2$ solution. After CsI was fully dissolved, 0.023 mg of $SnF_2$ was added to the solution to form $CsSnI_3$ perovskite precursor. For the film preparation, ITO glasses were sequentially cleaned in detergent, deionized water, acetone, isopropanol, and alcohol for 30 minutes. Afterward, the glasses were treated with ultraviolet ozone for 30 minutes. PEDOT: PSS films were coated on the cleaned ITO glasses at 6000 rpm for 40 s and then annealed in air at 140 °C for 20 min. The substrates were then transferred into a nitrogen-filled glovebox for deposition of the perovskite layer. Constant purging of the glovebox was carried out during the whole fabrication processes to maintain an oxygen level below 0.1 ppm. To deposit the perovskite film, the perovskite precursor (60 μl) was spun onto the PEDOT:PSS layer at 5000 rpm for 35 s with acceleration of 1000 rpm/s, Then the substrates were annealed at 100 °C for 10 min.

**$Cs_2AgBiBr_6$**

CsBr (170.25 mg, 0.8 mmol), AgBr (75.11 mg, 0.4 mmol), and $BiBr_3$ (179.48 mg, 0.4 mmol) were dissolved in DMSO (1 mL) to prepare a 0.4 M solution. The solution was stirred at 60 °C to ensure complete dissolution of the powders, and HBr (20 μL) was added to improve coverage. Before use, the solution was filtered with polytetrafluoroethylene (PTFE) syringe filters (0.22 μm). The glass substrates (1.5 $cm^2$) were cleaned sequentially by sonication in isopropanol, acetone, and isopropanol for 20 min each. The substrate surface was further treated with UV-Ozone for 15 min before use. 50 μL of the $Cs_2AgBiBr_6$ solution was dispersed onto the substrate, and the solution was initially spin-coated at 6000 rpm for 40 sec. The samples were annealed at 250 °C for 10 min to provide a light-yellow film. All the procedures were carried out inside the glovebox.

**Raman scattering spectroscopy**

Raman spectra were collected using the Edinburgh instrument RM5 Raman microscope stage. 785 nm laser excitation at 100× magnification was used. Samples were placed on a Linkam liquid nitrogen controlled microscopy stage installed on the Raman microscope stage and cooled down to 80 K. The stage was cooled to 313 K and purged with nitrogen before cooling to 80 K for these measurements. $CsPbI_3$ samples were annealed at 473 K, then cooled to 80 K to maintain their black phase. Once stabilized at 80K, the microscope was focused manually, and the Raman scattering spectra were measured with ten accumulations with a 3 s exposure time.

**Low-temperature photoluminescence studies**

To measure the PL spectra of the $CsPbI_3$ perovskite, a 450 nm laser was used to excite the material, producing a PL peak. For temperature-dependent measurements and to preserve the thin film phase, the sample was placed within a Linkam liquid nitrogen-controlled microscopy stage, and the laser beam was directed vertically through the sample window onto the thin film. The resulting PL emission was recorded after passing through a 500-nm long-pass glass filter to prevent detector oversaturation. After the samples were placed on the Linkam stage, they were purged with Nitrogen for 1 min and cooled to 80 K. The samples were heated by 5-10 K increments up to around 300 K, and their PL emission was recorded at each temperature step 1 min after temperature stabilization.

**Supporting data**

**Dual orbital model**

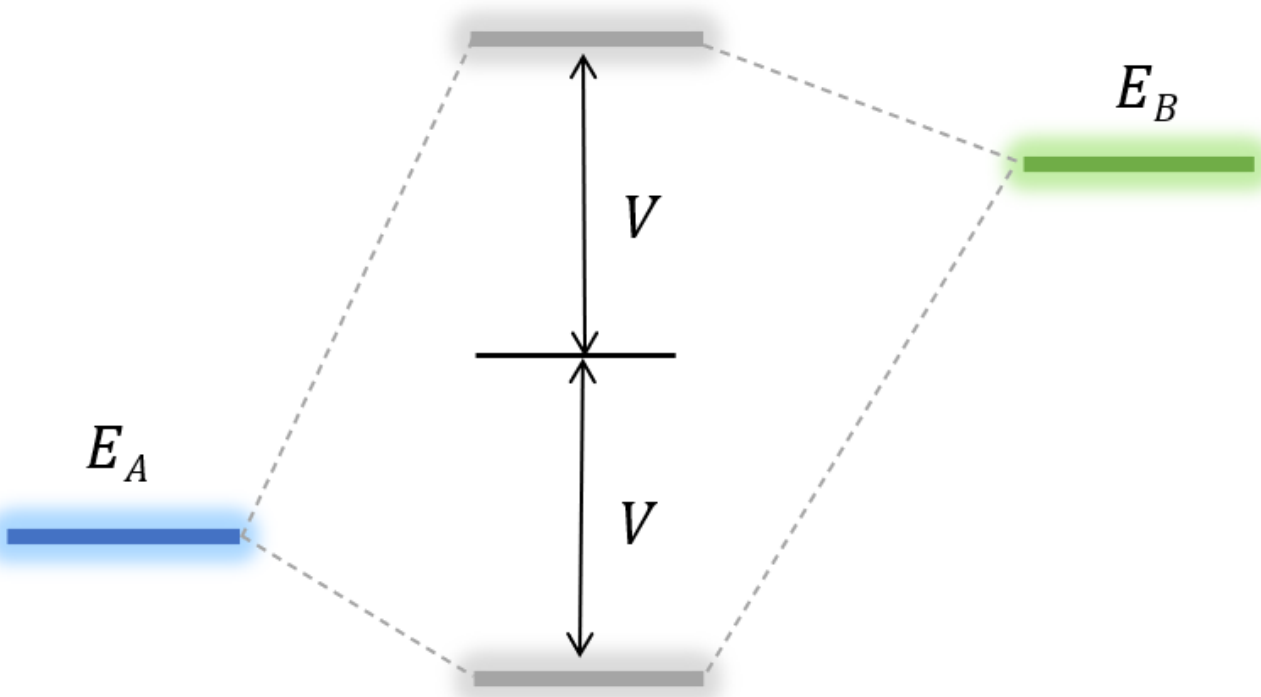

**Fig. S1 | A scheme of dual orbital model of linear combination atomic orbital.**

Here we first considered a simple dual orbital model, as shown in Fig. S1. The $E_A$, $E_B$ and $V$ are metal energy level A, halide energy level B and matrix element, respectively. Therefore, the *Hamiltonian* can be written as

$$H = \begin{pmatrix} E_{\mathrm{A}} & V \\ V & E_{\mathrm{B}} \end{pmatrix}$$

Here we define

$$\Delta E = |E_{\mathrm{A}} - E_{\mathrm{B}}|$$

The $V$ and $\Delta E$ describe the hybridization strength and energy difference between two orbitals, respectively. The energy decreasing can be written as $\Delta E_{bond} = -\frac{2V^2}{\Delta E}$. The $\Delta E_{bond}$ is direct related to ion migration energy barrier(*17*).

Note that the dual-orbital model does not aim to reproduce the full momentum- and band-resolved electron–phonon matrix elements $g_{mnv}(k, q)$ calculated by EPW. Instead, it provides an effective microscopic description of how a phonon-induced modulation of orbital hybridization governs the strength of electron–phonon interactions.

When phonons are taken into account, we have

$$V = V_0 + \frac{\partial V}{\partial Q_v} Q_v$$

Then the electron-phonon coupling comes from

$$g_{eff} = \left\langle A \middle| \frac{\partial H}{\partial Q_v} \middle| B \right\rangle$$

In then, the $g_{eff}$ can be written as

$$g_{eff} = \frac{V}{\sqrt{\Delta E^2 + 4V^2}} \frac{\partial V}{\partial d_{AB}}$$

where $d_{AB}$ is the distance between atom A and B. Further simplification under the weak hybridization limit yields

$$|g|^2 = \frac{V^2}{\Delta E^2}$$

Therefore, we can define $R = \frac{2V}{\Delta E}$ as a descriptor.

This descriptor is not merely the shared component of the two, but rather the tangent of the twice orbital mixing angle in a physically meaningful sense.

We now return to the Hamiltonian and express the bonding state as

$$|\psi\rangle = \cos\theta\,|A\rangle + \sin\theta|B\rangle$$

where θ is the orbital mixing angle. The rotation matrix is

$$U = \begin{pmatrix} \cos\theta & -\sin\theta \\ \sin\theta & \cos\theta \end{pmatrix}.$$

We require that $U^T HU$ be diagonal. In other words, the off-diagonal elements of the transformed matrix must vanish (i.e., be equal to zero). From the calculation we obtain

$$H'_{12} = \frac{\Delta E}{2} \sin 2\theta - V \cos 2\theta.$$

In order to achieve complete diagonalization, it is necessary that the new off-diagonal element vanishes, i.e. $H'_{12} = 0$. Then, we obtain

$$\frac{\Delta E}{2}\sin 2\theta = V\cos 2\theta.$$

That is

$$\tan 2\theta = \frac{2V}{\Delta E}$$

For diatomic systems, in general, hybridization is not limited to only two orbitals between the two atoms. Therefore, we need to consider the weights of additional orbital hybridizations, and these weights $c_i$ actually satisfy the hybridization state as following

$$c_i = \frac{(\frac{V_i}{\Delta E_i})^2}{\sum_i (\frac{V_i}{\Delta E_i})^2}$$

Therefore, the effective R between two atoms can be written as following

$$R = \sum\nolimits_i c_i R_i = \frac{\sum_i R_i^3}{\sum_i R_i^2}$$

**Crystal Structures**

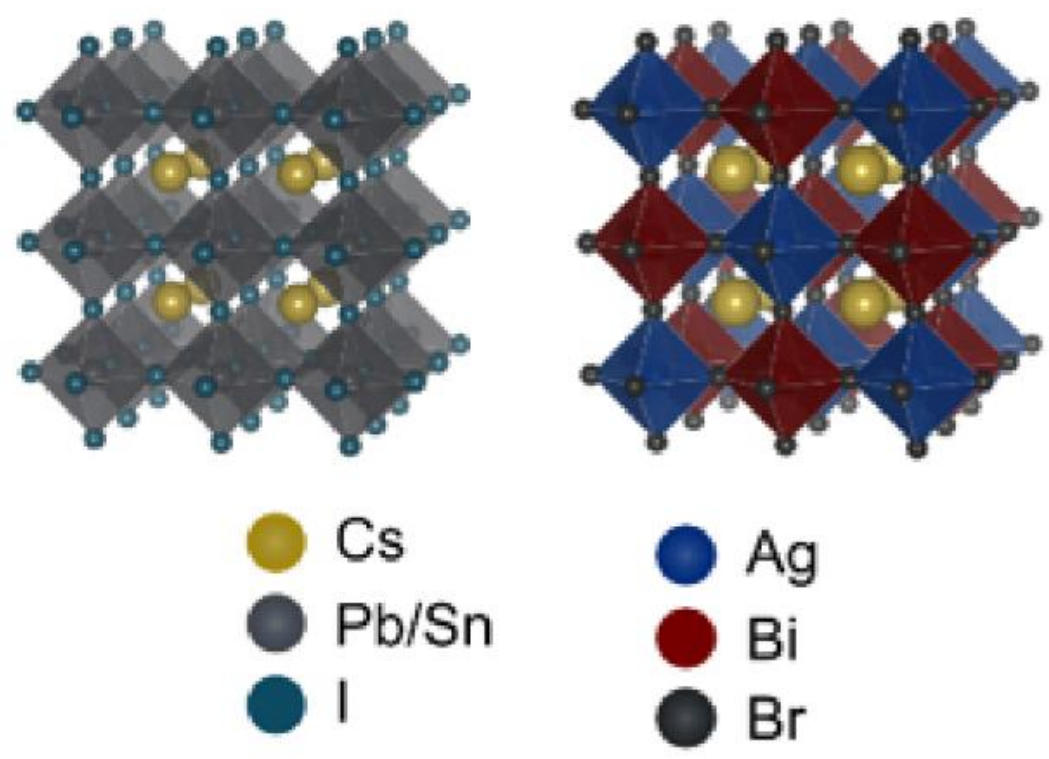

**Fig. S2 | Crystal structure for single and double halide perovskites.**

## Visualization of the Wannier orbital interpolation

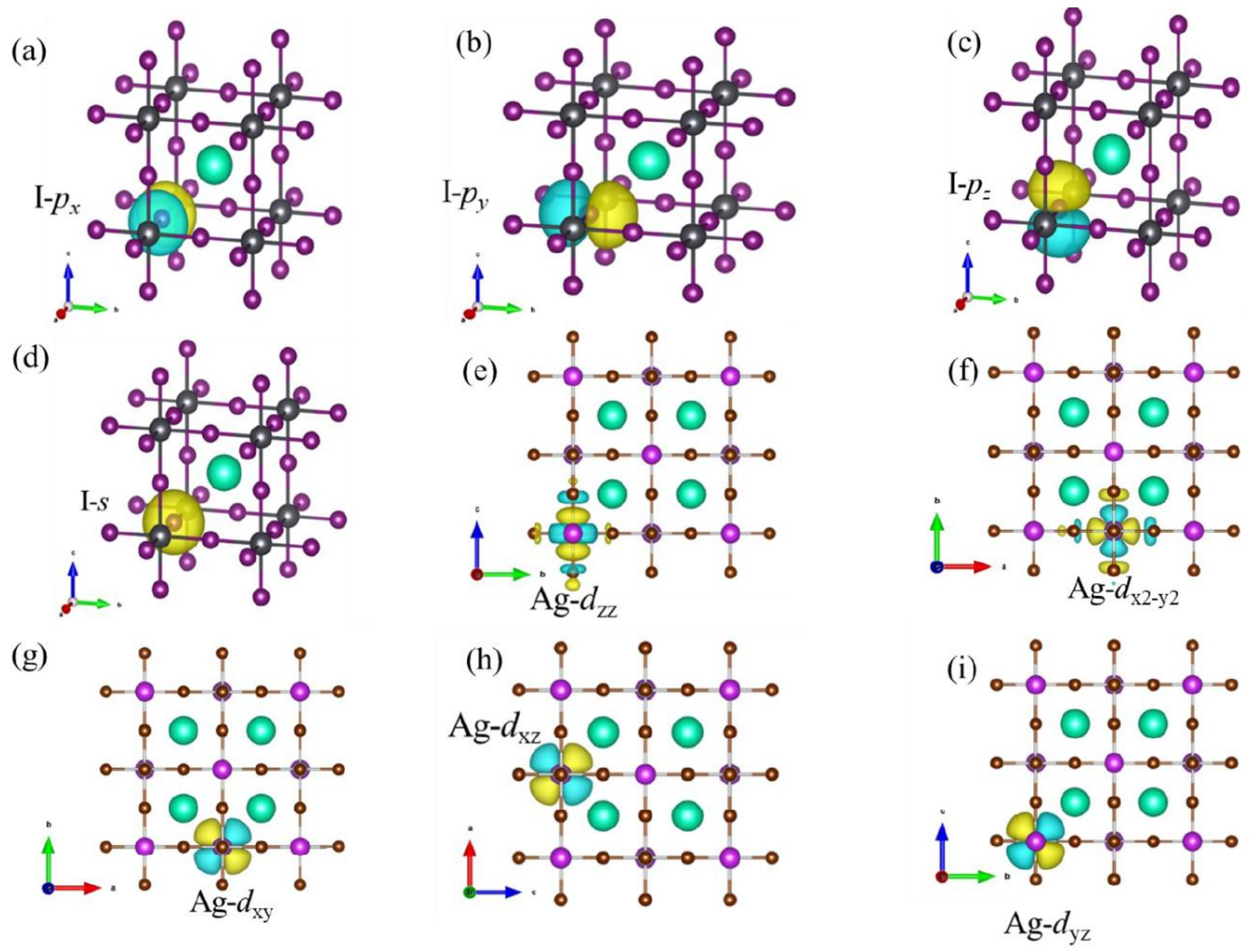

**Fig. S3 | Real-space isosurfaces of representative maximally localized Wannier functions (MLWFs) for cubic halide perovskites. a–d,** Iodine-derived I $p_x$, I $p_y$, I $p_z$ and I $s$ Wannier functions. **e–i**, Silver-derived Ag $d_{z^2}$, Ag $d_{x^2-y^2}$, Ag $d_{xy}$, Ag $d_{xz}$ and Ag $d_{yz}$ Wannier functions.

To efficiently interpolate the electronic structure, dynamical matrices and EPC matrix elements on dense k- and q-point grids, MLWFs were constructed for all three perovskite compositions. Since the calculations of the EPC strength, electron self-energy, and electron linewidth are all based on the Wannier representation, it is essential to verify that the constructed MLWFs possess well-defined orbital characters, sufficient spatial localization and high interpolation accuracy.

For $CsPbI_3$ and $CsSnI_3$, the highest valence band maximum (HVB) is primarily composed of I-5p states, whereas the lowest conduction band (LCB) mainly originates from Pb-6*p* and Sn-5*p* states, respectively. Accordingly, halogen *p* orbitals together with the B-site metal *p* orbitals were chosen as the initial projections to ensure an accurate description of the band-edge electronic states. In contrast, the electronic structure of $Cs_2AgBiBr_6$ is more complex because both the valence- and conduction-band edges involve multiple atomic species. Therefore, Br-*p* orbitals together with the relevant Ag- and Bi-derived orbitals were included in the Wannier projection.

The calculated Wannier spreads indicate that all MLWFs remain well localized in real space. The average spread of the MLWFs is approximately 2.22 Å$^2$ for $CsPbI_3$ and $CsSnI_3$, whereas the Ag-derived Wannier functions in $Cs_2AgBiBr_6$ exhibit a smaller average spread of approximately 1.62 Å$^2$. This enhanced localization reflects the stronger orbital confinement introduced by the heterovalent double-B-site framework. Consequently, the electronic wavefunctions become less delocalized, leading to a more localized polarization response that provides a favorable electronic-structure basis for stronger long-range Fröhlich electron–phonon coupling. The different degrees of Wannier localization therefore not only determine the accuracy of the interpolation but also reflect intrinsic differences in orbital hybridization and polarization response among the three perovskite systems.

Fig. S3 presents representative real-space isosurfaces of the constructed MLWFs for $CsPbI_3$, $CsSnI_3$, and $Cs_2AgBiBr_6$, including the halogen *s* and *p* orbitals together with the relevant metal-centered orbitals. All Wannier functions are well localized around their corresponding atomic sites and preserve the expected orbital symmetries. For example, the halogen *p* orbitals exhibit the characteristic orientations of the $p_x$, $p_y$ and $p_z$ orbitals, whereas the Ag-derived Wannier functions clearly reproduce the five *d*-orbital symmetries $d_{z^2}$, $d_{x^2-y^2}$, $d_{xy}$, $d_{xz}$ and $d_{yz}$. These results demonstrate that the Wannierization faithfully preserves the underlying orbital characters while accurately describing the local bonding environments in each composition. In particular, the more intricate spatial distributions of the Ag- and Bi-derived Wannier functions in $Cs_2AgBiBr_6$ reflect the pronounced reconstruction of the local electronic structure induced by the heterovalent double-B-site ordering. This increased electronic-structure complexity enhances charge-density inhomogeneity and local polarization, providing a microscopic origin for the stronger Fröhlich electron–phonon coupling observed in the double perovskite.

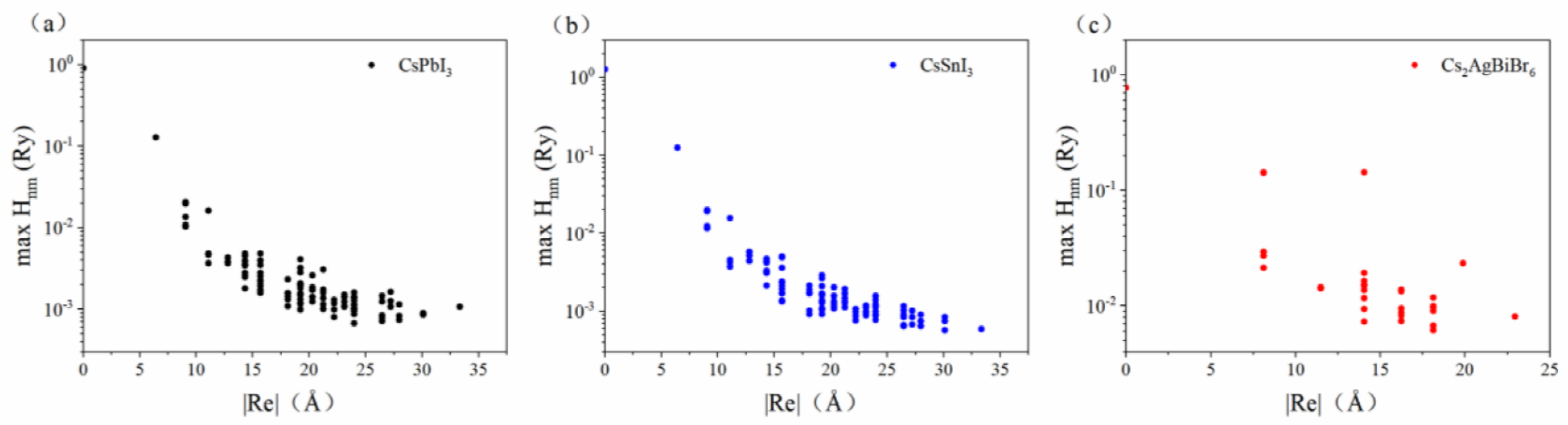

**Fig. S4 | Spatial decay of the real-space Wannier Hamiltonian for cubic halide perovskites.** **a,** $CsPbI_3$. **b,** $CsSnI_3$. **c,** $Cs_2AgBiBr_6$.

In addition to the spatial localization of the MLWFs, the accuracy of Wannier interpolation also depends critically on the locality of the real-space Hamiltonian. Fig. S4 shows the decay of the Hamiltonian matrix elements, H(R), in the Wannier representation as a function of the distance between Wannier centers. For all three perovskite compositions, the Hamiltonian matrix elements decay rapidly with increasing separation and become negligible beyond a short real-space distance, indicating that the electronic interactions are predominantly short ranged in the Wannier basis. This rapid decay demonstrates the excellent locality of the Wannier Hamiltonian and confirms that the electronic structure can be accurately reconstructed using a finite real-space cutoff. Consequently, the Wannier representation provides an efficient and reliable framework for interpolating the electronic band structure on dense *k*-point grids while retaining first-principles accuracy. The strong locality of the Wannier Hamiltonian therefore provides the foundation for accurate interpolation of the electronic structure, electron velocities, and electron–phonon coupling matrix elements on dense Brillouin-zone grids.

## HSE+SOC band structures

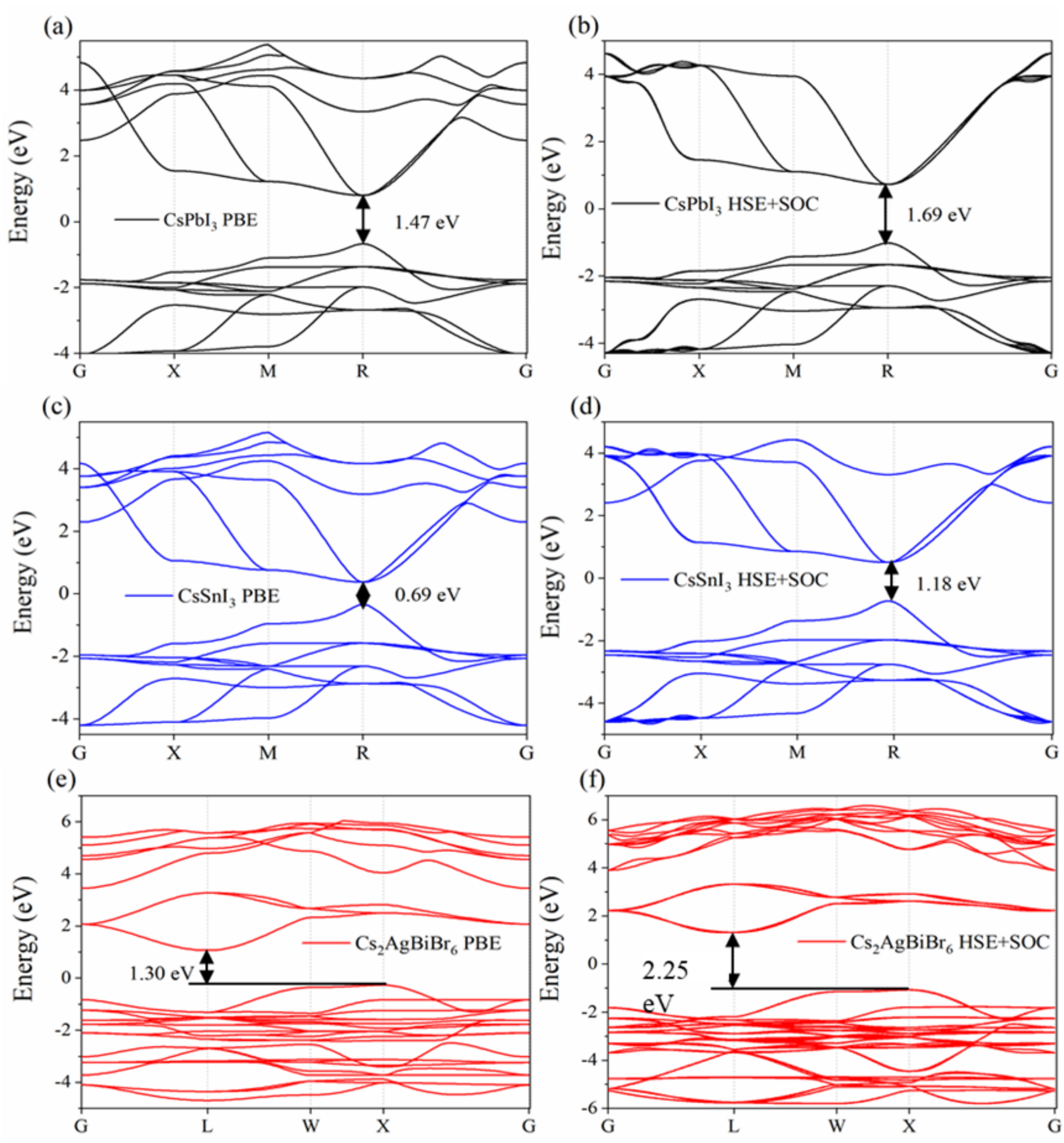

**Fig. S5 | Comparison of PBE band structures and HSE+SOC band structures.**

## Projected Crystal Orbital Hamilton Population

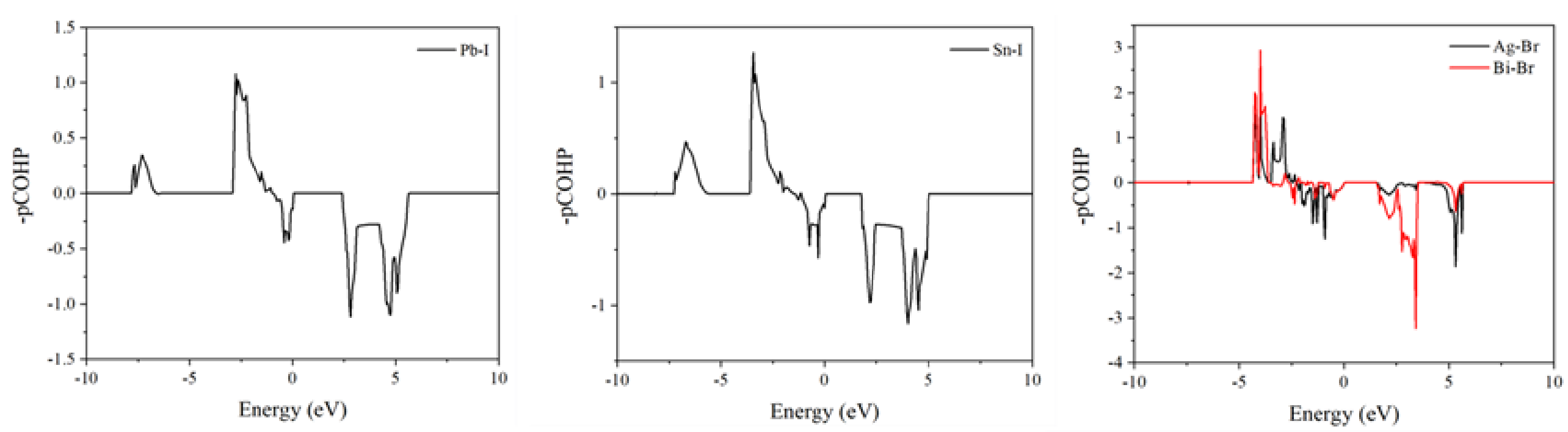

**Fig. S6 | The Projected Crystal Orbital Hamilton Population (pCOHP) of a, $CsPbI_3$ b, $CsSnI_3$ and c, $Cs_2AgBiBr_6$.** The results indicate that the highest valence bands for three halide perovskites are dominated by antibonding states of B-site cations and X-site anions.

## Static descriptor R in three halide perovskites

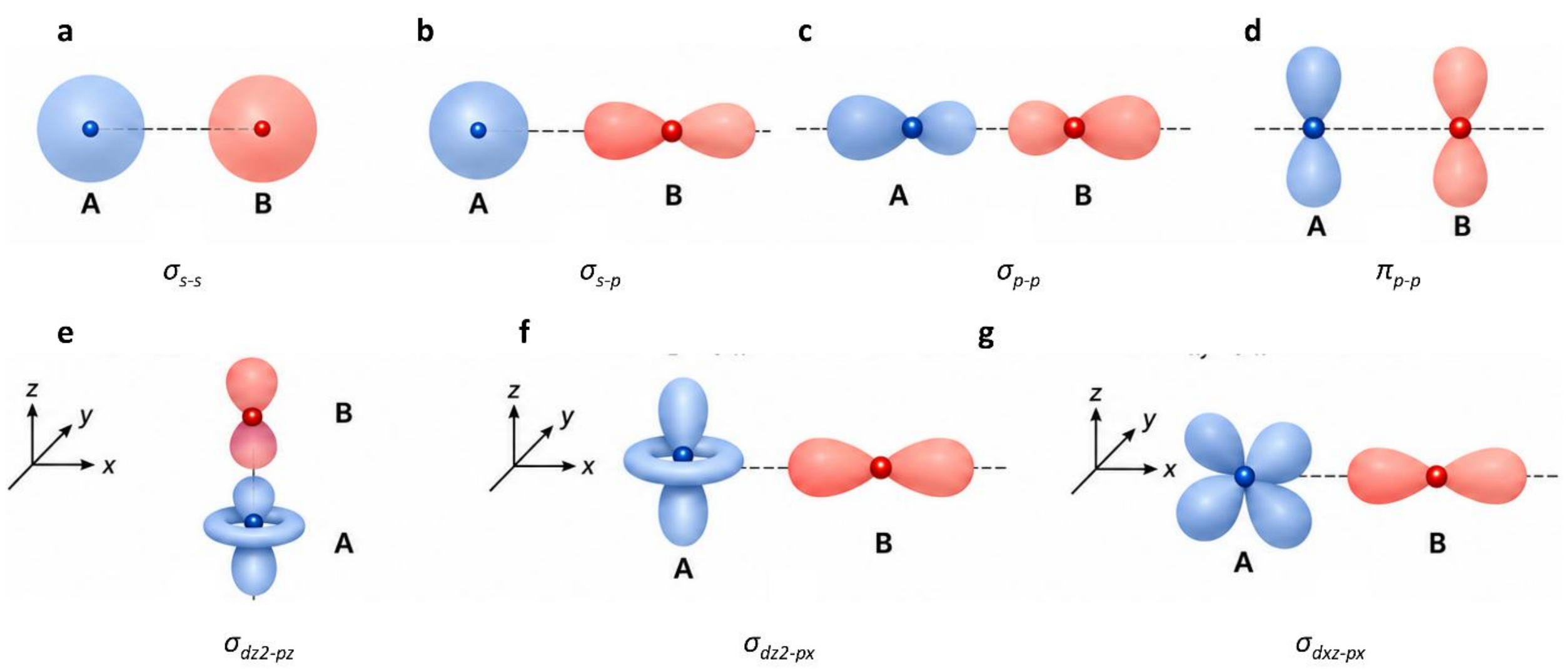

**Fig. S7 | Selected orbitals and their spatial arrangement for extracting R. a,** $\sigma_{s-s}$, **b,** $\sigma_{s-p}$, **c,** $\sigma_{p-p}$, **d,** $\pi_{p-p}$, **e,** $\sigma_{d_{z^2}-p_z}$, **f,** $\sigma_{d_{z^2}-p_x}$, **g,** $\sigma_{d_{xz}-p_z}$.

**Table S1 | Orbital hybridization states of $CsPbI_3$ at fully-relaxed lattice, the corresponding arrangement is shown in Fig. S7. The effective R between two atoms is 0.48.**

| orbital i | orbital j | arrangement | ΔE | V | R |
|---|---|---|---|---|---|
| Pb s | I s | a | 5.34 | 0.35 | 0.13 |
| Pb s | I p | b | 3.90 | 1.09 | 0.56 |
| Pb p | I p | c | 5.49 | 0.71 | 0.26 |
| Pb p | I p | d | 4.30 | 0.24 | 0.11 |

**Table S2 | Orbital hybridization states of $CsSnI_3$ at fully-relaxed lattice, the corresponding arrangement is shown in Fig. S7. The effective R between two atoms is 0.71.**

| orbital i | orbital j | arrangement | ΔE | V | R |
|---|---|---|---|---|---|
| Sn s | I s | a | 6.43 | 0.37 | 0.11 |
| Sn s | I p | b | 2.88 | 1.13 | 0.79 |
| Sn p | I p | c | 5.42 | 0.67 | 0.25 |
| Sn p | I p | d | 3.73 | 0.30 | 0.16 |

**Table S3 | Orbital hybridization states of $Cs_2AgBiBr_6$ at fully-relaxed lattice, the corresponding arrangement is shown in Fig. S7. The effective R between Ag-Br and Bi-Br are 0.57 and 1.69, respectively.**

| orbital i | orbital j | arrangement | ΔE | V | R |
|---|---|---|---|---|---|
| Ag $e_g$ | Br p | e | 4.13 | 0.75 | 0.36 |
| Ag $e_g$ | Br p | f | 4.12 | 0.43 | 0.21 |
| Ag $t_{2g}$ | Br p | g | 0.90 | 0.30 | 0.67 |
| Bi s | Br s | a | 1.08 | 1.01 | 1.87 |
| Bi s | Br p | b | 3.58 | 0.32 | 0.18 |
| Bi p | Br s | b | 9.52 | 2.05 | 0.43 |
| Bi p | Br p | c | 7.02 | 0.85 | 0.24 |
| Bi p | Br p | d | 2.16 | 0.51 | 0.47 |

**Dynamical descriptor R(δ) in the vibrational-perturbed structure**

Since both ion migration and electron–phonon coupling actually involve atomic displacements, we here perturb $R$ with respect to the displacement $\delta$

$$V = V(\delta)$$

$$\Delta E = \Delta E(\delta)$$

$$R(\delta) = \frac{2V(\delta)}{\Delta E(\delta)}$$

expanding the equation

$$R(\delta) = R_0 + \frac{\partial R}{\partial \delta}\delta + \cdots$$

Then, we can obtain orbital hybridization susceptibility $\chi_R$

$$\chi_R = \frac{\partial R}{\partial \delta}$$

Regarding the EPC,

$$g_{mn\nu} = <m \mid \frac{\partial V_{KS}}{\partial Q_\nu} \mid n>.$$

Since the Hamiltonian depends on $R$, we have $H = H(R)$.

For a general lattice vibration, the atomic displacements $\delta$ can be expressed in terms of the phonon normal coordinate,

$$\delta_i = \frac{e_{i\nu}}{\sqrt{M_i}} Q_\nu.$$

Applying the chain rule gives

$$\frac{\partial H}{\partial Q} = \frac{\partial H}{\partial R}\frac{\partial R}{\partial Q}.$$

Then, $g$ is approximately proportional to $\partial R/\partial Q$

$$g \propto \frac{\partial R}{\partial Q}.$$

This explains why stronger lattice-induced changes in orbital hybridization lead to enhanced electron–phonon coupling.

The static R determines the intrinsic properties of the material, while the dynamic R($\delta$) reflects the modulation of these properties by different lattice vibrations.

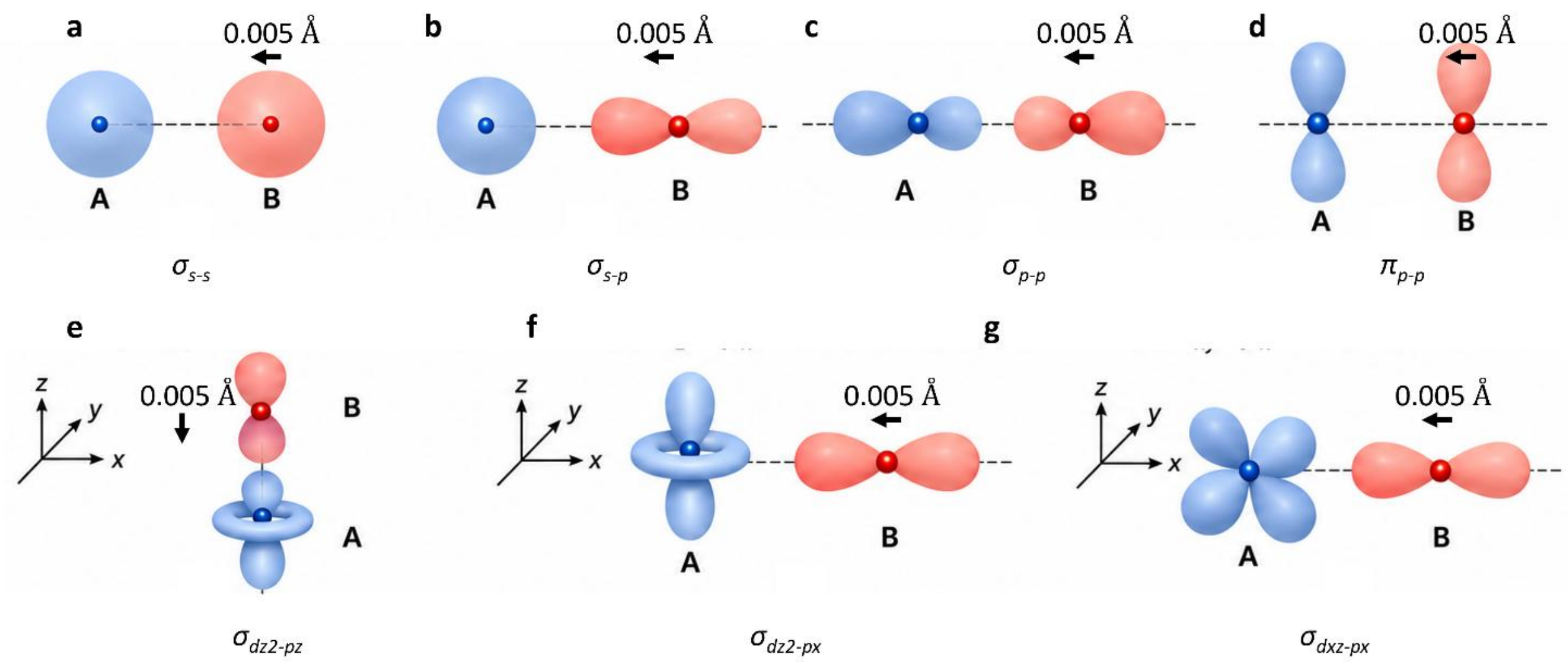

**Fig. S8 | Selected orbitals, their spatial arrangement and vibrational-perturbed direction (stretching modes, slightly shortened bond lengths) for extracting R. a,** $\sigma_{s-s}$, **b,** $\sigma_{s-p}$, **c,** $\sigma_{p-p}$, **d,** $\pi_{p-p}$, **e,** $\sigma_{d_{z^2}-p_z}$, **f,** $\sigma_{d_{z^2}-p_x}$, **g,** $\sigma_{d_{xz}-p_z}$.

**Table S4 | Orbital hybridization states of $CsPbI_3$ at slightly shortened bond lengths, the corresponding arrangement is shown in Fig. S8. The effective R between two atoms is 0.57.**

| orbital i | orbital j | arrangement | ΔE($\delta$) | V($\delta$) | R($\delta$) |
|---|---|---|---|---|---|
| Pb s | I s | a | 5.20 | 0.33 | 0.13 |
| Pb s | I p | b | 3.95 | 0.70 | 0.35 |
| Pb p | I p | c | 4.07 | 1.33 | 0.66 |
| Pb p | I p | d | 4.43 | 0.26 | 0.12 |

**Table S5 | Orbital hybridization states of $CsSnI_3$ at slightly shortened bond lengths, the corresponding arrangement is shown in Fig. S8. The effective R between two atoms is 0.56.**

| orbital i | orbital j | arrangement | $\Delta E(\delta)$ | $V(\delta)$ | $R(\delta)$ |
|---|---|---|---|---|---|
| Sn s | I s | a | 6.51 | 0.39 | 0.12 |
| Sn s | I p | b | 2.89 | 0.82 | 0.57 |
| Sn p | I p | c | 4.06 | 1.23 | 0.60 |
| Sn p | I p | d | 3.67 | 0.31 | 0.17 |

**Table S6 | Orbital hybridization states of $Cs_2AgBiBr_6$ at slightly shortened bond lengths, the corresponding arrangement is shown in Fig. S8. The effective R between Ag-Br and Bi-Br are 0.58 and 1.61, respectively.**

| orbital i | orbital j | arrangement | $\Delta E(\delta)$ | $V(\delta)$ | $R(\delta)$ |
|---|---|---|---|---|---|
| Ag $e_g$ | Br p | e | 4.13 | 0.75 | 0.36 |
| Ag $e_g$ | Br p | f | 4.13 | 0.43 | 0.21 |
| Ag $t_{2g}$ | Br p | g | 0.90 | 0.30 | 0.68 |
| Bi s | Br s | a | 0.67 | 0.60 | 1.81 |
| Bi s | Br p | b | 3.17 | 0.15 | 0.09 |
| Bi p | Br s | b | 9.08 | 2.25 | 0.49 |

| Bi p | Br p | c | 6.57 | 0.91 | 0.28 |
|---|---|---|---|---|---|
| Bi p | Br p | d | 2.18 | 0.52 | 0.47 |

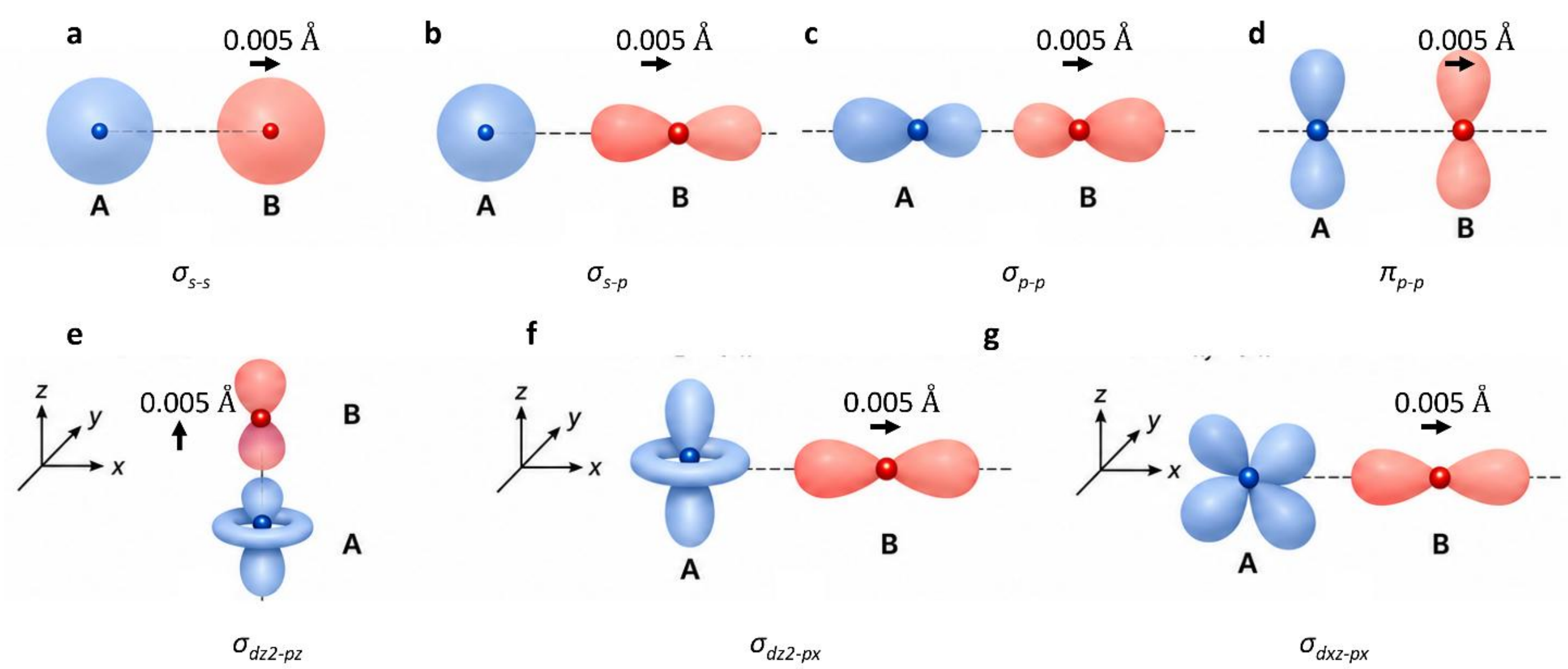

**Fig. S9 | Selected orbitals, their spatial arrangement and vibrational-perturbed direction (stretching mode, slightly elongated bond lengths) for extracting R. a,** $\sigma_{s-s}$, **b,** $\sigma_{s-p}$, **c,** $\sigma_{p-p}$, **d,** $\pi_{p-p}$, **e,** $\sigma_{d_{z^2}-p_z}$, **f,** $\sigma_{d_{z^2}-p_x}$, **g,** $\sigma_{d_{xz}-p_z}$.

**Table S7 | Orbital hybridization states of $CsPbI_3$ at slightly elongated bond lengths, the corresponding arrangement is shown in Fig. S9. The effective R between two atoms is 0.67.**

| orbital i | orbital j | arrangement | ΔE($\delta$) | V($\delta$) | R($\delta$) |
|---|---|---|---|---|---|
| Pb s | I s | a | 5.20 | 0.22 | 0.08 |
| Pb s | I p | b | 3.93 | 1.46 | 0.75 |
| Pb p | I p | c | 4.09 | 0.61 | 0.30 |
| Pb p | I p | d | 4.40 | 0.25 | 0.11 |

**Table S8 | Orbital hybridization states of $CsSnI_3$ at slightly elongated bond lengths, the corresponding arrangement is shown in Fig. S9. The effective R between two atoms is 0.74.**

| orbital i | orbital j | arrangement | ΔE($\delta$) | V($\delta$) | R($\delta$) |
|---|---|---|---|---|---|
| Sn s | I s | a | 6.49 | 0.39 | 0.12 |
| Sn s | I p | b | 2.90 | 1.25 | 0.86 |
| Sn p | I p | c | 4.05 | 0.94 | 0.46 |
| Sn p | I p | d | 3.64 | 0.30 | 0.16 |

**Table S9 | Orbital hybridization states of $Cs_2AgBiBr_6$ at slightly elongated bond lengths, the corresponding arrangement is shown in Fig. S9. The effective R between Ag-Br and Bi-Br are 0.56 and 3.83, respectively.**

| orbital i | orbital j | arrangement | ΔE($\delta$) | V($\delta$) | R($\delta$) |
|---|---|---|---|---|---|
| Ag $e_g$ | Br p | e | 4.12 | 0.74 | 0.36 |
| Ag $e_g$ | Br p | f | 4.12 | 0.43 | 0.21 |
| Ag $t_{2g}$ | Br p | g | 0.90 | 0.30 | 0.66 |
| Bi s | Br s | a | 0.69 | 1.37 | 3.94 |
| Bi s | Br p | b | 3.21 | 0.47 | 0.29 |
| Bi p | Br s | b | 9.08 | 1.78 | 0.39 |
| Bi p | Br p | c | 6.57 | 0.76 | 0.23 |
| Bi p | Br p | d | 2.15 | 0.50 | 0.47 |

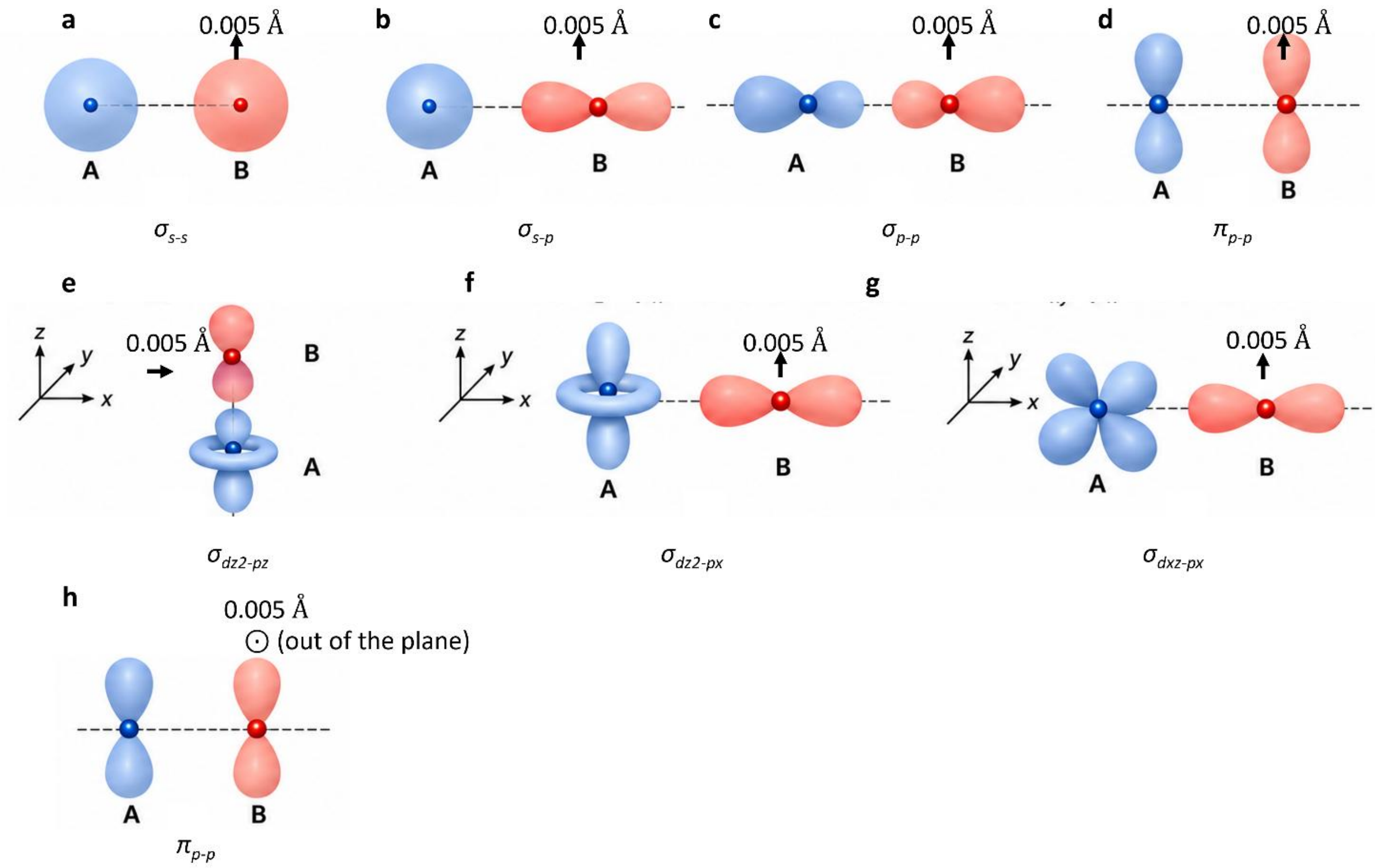

**Fig. S10 | Selected orbitals, their spatial arrangement and vibrational-perturbed direction (shearing mode, slightly shifted perpendicular to the bonding direction) for extracting R. a,** $\sigma_{s-s}$, **b,** $\sigma_{s-p}$, **c,** $\sigma_{p-p}$, **d,** $\pi_{p-p}$, **e,** $\sigma_{d_{z^2}-p_z}$, **f,** $\sigma_{d_{z^2}-p_x}$, **g,** $\sigma_{d_{xz}-p_z}$, **h,** $\pi_{p-p}$

**Table S10 | Orbital hybridization states of $CsPbI_3$ at slightly shifted perpendicular to the bonding direction, the corresponding arrangement is shown in Fig. S10. The effective R between two atoms is 0.45.**

| orbital i | orbital j | arrangement | ΔE($\delta$) | V($\delta$) | R($\delta$) |
|---|---|---|---|---|---|
| Pb s | I s | a | 5.34 | 0.35 | 0.13 |
| Pb s | I p | b | 3.91 | 1.09 | 0.56 |
| Pb p | I p | c | 5.36 | 0.78 | 0.29 |
| Pb p | I p | d | 3.29 | 0.30 | 0.18 |

| Pb p | I p | h | 4.27 | 0.25 | 0.12 |
|---|---|---|---|---|---|

**Table S11 | Orbital hybridization states of $CsSnI_3$ at slightly shifted perpendicular to the bonding direction, the corresponding arrangement is shown in Fig. S10. The effective R between two atoms is 0.61.**

| orbital i | orbital j | arrangement | ΔE($\delta$) | V($\delta$) | R($\delta$) |
|---|---|---|---|---|---|
| Sn s | I s | a | 6.51 | 0.38 | 0.12 |
| Sn s | I p | b | 2.78 | 1.03 | 0.74 |
| Sn p | I p | c | 4.46 | 0.95 | 0.43 |
| Sn p | I p | d | 3.44 | 0.33 | 0.19 |
| Sn p | I p | h | 3.55 | 0.30 | 0.17 |

**Table S12 | Orbital hybridization states of $Cs_2AgBiBr_6$ at slightly shifted perpendicular to the bonding direction, the corresponding arrangement is shown in Fig. S10. The effective R between Ag-Br and Bi-Br are 0.57 and 2.67, respectively.**

| orbital i | orbital j | arrangement | ΔE($\delta$) | V($\delta$) | R($\delta$) |
|---|---|---|---|---|---|
| Ag $e_g$ | Br p | e | 4.13 | 0.75 | 0.36 |
| Ag $e_g$ | Br p | f | 4.13 | 0.43 | 0.21 |
| Ag $t_{2g}$ | Br p | g | 0.90 | 0.30 | 0.67 |
| Bi s | Br s | a | 0.68 | 0.98 | 2.89 |
| Bi s | Br p | b | 3.18 | 0.31 | 0.19 |
| Bi p | Br s | b | 9.52 | 2.06 | 0.43 |
| Bi p | Br p | c | 7.02 | 0.85 | 0.24 |
| Bi p | Br p | d | 1.71 | 0.50 | 0.59 |
| Bi p | Br p | h | 2.16 | 0.51 | 0.47 |

**Ion migration pathway diagram**

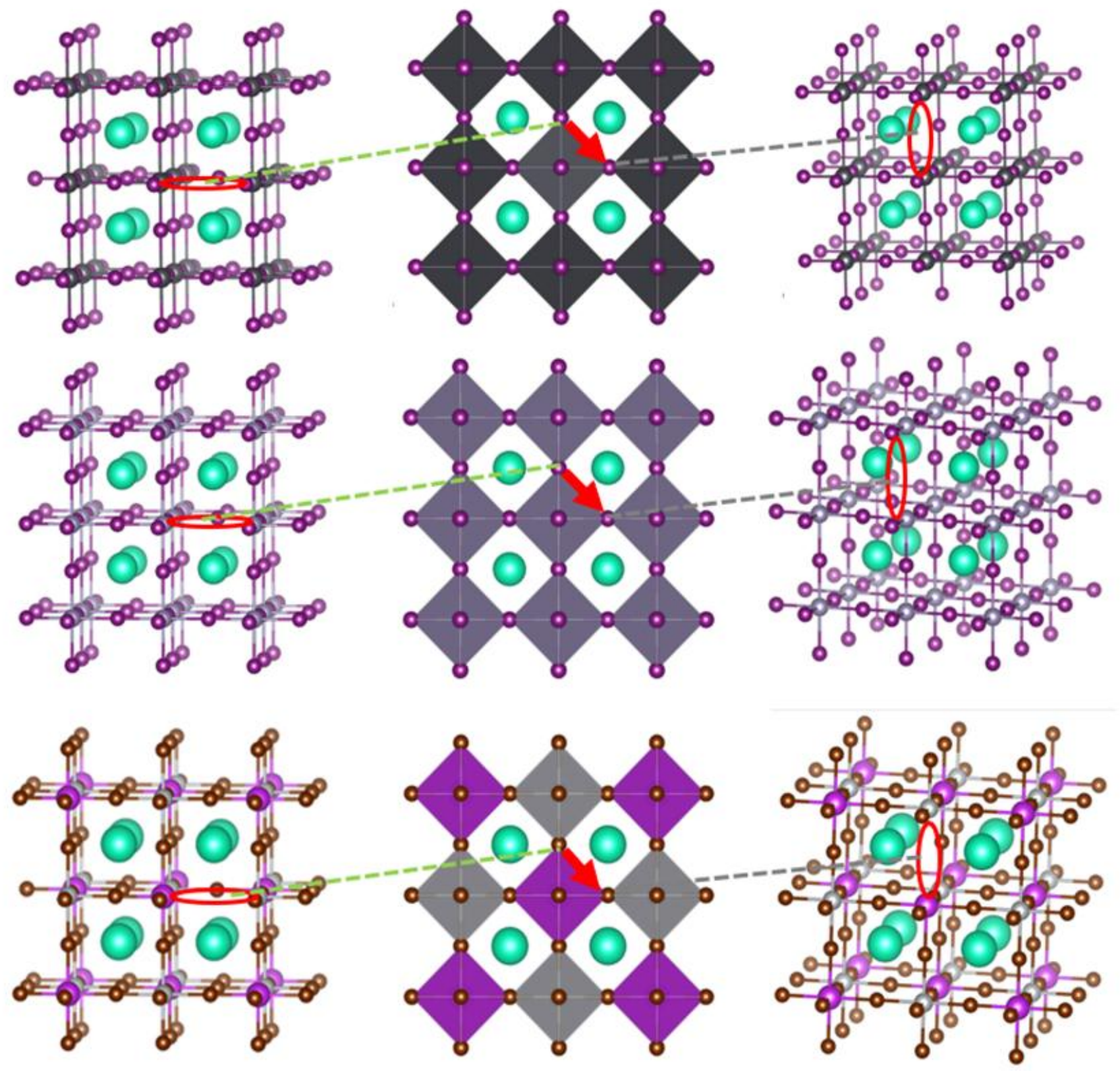

**Fig. S11 | The potential ion migration pathway of X-site anions.**

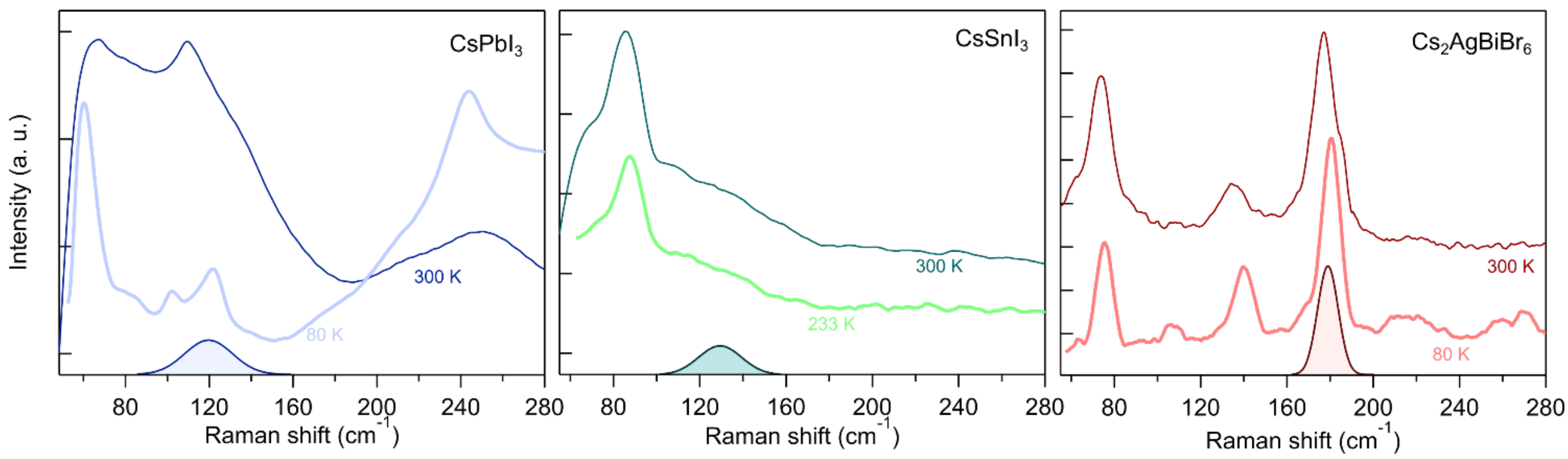

**Fig. S12 | Thermal stability of LO phonons at low temperature.** A comparison of the Raman scattering spectra recorded from the three perovskite compositions at low temperatures, using a liquid nitrogen cryogenic system and 785 nm excitation. The spectra reflect their Raman active modes at the zone center (Γ point). Taking the LO phonon position as the high-frequency feature in the optical bands (by the solid peak lineshapes)., we determine phonon energies of $CsSnI_3$, $CsPbI_3$, and $Cs_2AgBiBr_6$ to be respectively 14.5, 14.9, and 21.7 meV. Only minor changes in the energy are seen with varied temperature, indicating they are thermally stable. Note that the lowest temperature that the Raman spectra could be reasonably resolved in $CsSnI_3$ was near 233 K, as the Raman excitation line (1.58 eV) is in close proximity to the growing luminescence background generated in this narrow bandgap material, drowning out the Raman scattering signal when cooled to lower temperatures.

## References

1. Giannozzi, P. *et al.* Advanced capabilities for materials modelling with QUANTUM ESPRESSO. *J. Phys. Condens. Matter* **29**, 465901 (2017).

2. Giannozzi, P. *et al.* QUANTUM ESPRESSO: a modular and open-source software project for quantum simulations of materials. *J. Phys. Condens. Matter* **21**, 395502 (2009).

3. Blöchl, P. E. Projector augmented-wave method. *Phys. Rev. B* **50**, 17953–17979 (1994).

4. Hamann, D. R., Schlüter, M. & Chiang, C. Norm-Conserving Pseudopotentials. *Phys. Rev. Lett.* **43**, 1494–1497 (1979).

5. Monkhorst, H. J. & Pack, J. D. Special points for Brillouin-zone integrations. *Phys. Rev. B* **13**, 5188–5192 (1976).

6. Perdew, J. P., Burke, K. & Ernzerhof, M. Generalized Gradient Approximation Made Simple. *Phys. Rev. Lett.* **77**, 3865–3868 (1996).

7. Heyd, J., Scuseria, G. E. & Ernzerhof, M. Hybrid functionals based on a screened Coulomb potential. *J. Chem. Phys.* **118**, 8207–8215 (2003).

8. Zacharias, M., Volonakis, G., Giustino, F. & Even, J. Anharmonic lattice dynamics via the special displacement method. *Phys. Rev. B* **108**, 035155 (2023).

9. Henkelman, G., Uberuaga, B. P. & Jónsson, H. A climbing image nudged elastic band method for finding saddle points and minimum energy paths. *J. Chem. Phys.* **113**, 9901–9904 (2000).

10. Deringer, V. L., Tchougréeff, A. L. & Dronskowski, R. Crystal Orbital Hamilton Population (COHP) Analysis As Projected from Plane-Wave Basis Sets. *J. Phys. Chem. A* **115**, 5461–5466

(2011).

11. Dronskowski, R. & Bloechl, P. E. Crystal orbital Hamilton populations (COHP): energy-resolved visualization of chemical bonding in solids based on density-functional calculations. *J. Phys. Chem.* **97**, 8617–8624 (1993).

12. Maintz, S., Deringer, V. L., Tchougréeff, A. L. & Dronskowski, R. LOBSTER: A tool to extract chemical bonding from plane-wave based DFT. *J. Comput. Chem.* **37**, 1030–1035 (2016).

13. Lee, H. *et al.* Electron–phonon physics from first principles using the EPW code. *Npj Comput. Mater.* **9**, 156 (2023).

14. Noffsinger, J. *et al.* EPW: A program for calculating the electron–phonon coupling using maximally localized Wannier functions. *Comput. Phys. Commun.* **181**, 2140–2148 (2010).

15. Poncé, S., Margine, E. R., Verdi, C. & Giustino, F. EPW: Electron–phonon coupling, transport and superconducting properties using maximally localized Wannier functions. *Comput. Phys. Commun.* **209**, 116–133 (2016).

16. Cai, B. *et al.* A New Descriptor for Complicated Effects of Electronic Density of States on Ion Migration. *Adv. Funct. Mater.* **33**, 2300445 (2023).